\documentclass[journal=jctcce,manuscript=article]{achemso}
\setkeys{acs}{maxauthors=0,articletitle=true}
\usepackage{achemso}
\usepackage{placeins}
\usepackage{graphics}
\usepackage{amssymb,amsfonts}
\usepackage{graphicx}
\usepackage[table,dvipsnames]{xcolor}
\usepackage{multirow}
\usepackage{caption}
\usepackage{subcaption}
\usepackage{booktabs}
\usepackage{colortbl}
\usepackage{amsmath}
\usepackage{amsopn}
\usepackage{bm}
\usepackage{braket}
\usepackage{siunitx}
\usepackage{color}
\usepackage{array}
\usepackage{lscape}
\usepackage{mciteplus}
\usepackage[version=3]{mhchem}
\usepackage{ulem}
\usepackage{listings}
\usepackage{enumerate}
\usepackage{lmodern}
\usepackage{mhchem}
\usepackage[implicit=false]{hyperref}

\definecolor{green}{HTML}{66c2a5}
\definecolor{orange}{HTML}{fc8d62}

\author{Jonas Greiner}
\affiliation[mainz]{Department Chemie, Johannes Gutenberg-Universit{\"a}t Mainz\\Duesbergweg 10--14, 55128 Mainz, Germany}
\author{J{\"u}rgen Gauss}
\email{gauss@uni-mainz.de}
\affiliation[mainz]{Department Chemie, Johannes Gutenberg-Universit{\"a}t Mainz\\Duesbergweg 10--14, 55128 Mainz, Germany}
\author{Janus J. Eriksen}
\email{janus@dtu.dk}
\affiliation[dtu]{DTU Chemistry, Technical University of Denmark\\Kemitorvet Bldg. 206, 2800 Kgs. Lyngby, Denmark}

\title[TITLE]{Error Control and Automatic Detection of Reference Active Spaces in Many-Body Expanded Full Configuration Interaction}

\begin{document}

\begin{abstract}

We present a wide-reaching revamp of the generalized many-body expanded full configuration interaction (MBE-FCI) method. First, we outline how to automatize the selection of reference active spaces whereby the inherent bias introduced through a manual identification is reduced, also within the context of traditional complete active space methods. Second, we allow for the use of compact orbital clusters as expansion objects, which works to circumvent the unfavorable scaling with the number of orbitals included in the space complementary to the reference orbitals. Finally, we present a new algorithm for ensuring that many-body expansions can be efficiently terminated while conservatively accounting for resulting errors. These developments are all tested on a variety of molecular systems and different orbital representations to illustrate the abilities of our algorithm to produce correlation energies within predetermined error bounds, significantly broadening the overall applicability of the MBE-FCI method.

\end{abstract}

\newpage

\begin{figure}[ht]
\begin{center}
\includegraphics[width=\textwidth]{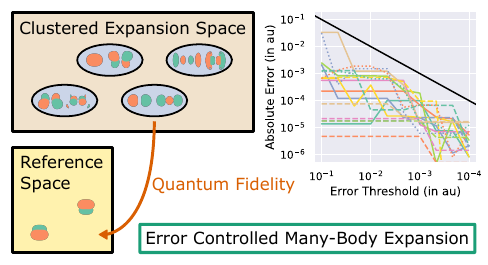}
\caption*{TOC graphic.}
\label{toc_fig}
\end{center}
\end{figure}

\newpage

\section{Introduction}\label{intro_sect}

In the description of the ground-state electronic structure of the overwhelming majority of all chemical compounds, the Hartree-Fock (HF) Slater determinant will constitute the dominant contribution to the exact full configuration interaction (FCI) wave function, as obtained by means of the variational principle~\cite{eriksen2021a}. However, for certain chemical systems, e.g., transient species, radicals, geometrically distorted molecules, as well as organometallic and a variety of inorganic compounds, the weight of the HF determinant in the linear FCI wave function may plummet significantly while those corresponding to other electronic configurations will grow accordingly. For quantum-chemical calculations of such inherently multireference systems, a careful consideration of the electronic structure within an active space, from which to spawn these additional determinants, is mandated, and it may even prove necessary to seek to relax the molecular orbitals (MOs) with respect to the electron correlations within this dedicated space over the canonical mean-field counterparts.\\

As an alternative to exact diagonalization or, more commonly, a Davidson procedure to obtain an FCI wave function in the complete space spanned by a particular set of MOs, the many-body expanded FCI (MBE-FCI) method approximates energies and properties within a given Hilbert space without recourse to an explicit sampling of the wave function~\cite{Eriksen2017,Eriksen2018,Eriksen2019,Eriksen2019a,eriksen2020,eriksen2021}. Instead, a multitude of complete active space CI (CASCI) calculations are performed on restricted MO tuples of increasing size, the results of which are incrementally recombined to yield total energies and properties. In the current formulation of the MBE-FCI method, correlation energies and first-order correlation properties are decomposed by first enforcing a strict partitioning of a complete set of MOs into a reference and an expansion space, upon which the residual correlation in the latter of these two spaces is recovered by means of an MBE in a given set of MOs:
\begin{align}
E - E^{\text{ref}} &= \sum_p\epsilon_p + \sum_{p<q}\Delta\epsilon_{pq} + \sum_{p<q<r}\Delta\epsilon_{pqr} + \ldots \label{mbe_eq}
\end{align}
In Eq. \ref{mbe_eq}, $E^{\text{ref}}$ describes the CASCI correlation energy of the reference space orbitals, while contributions to the full correlation energy from an MBE in the expansion space are expressed in terms of increments, $\Delta\epsilon$. The increments are recursively defined as in Ref.~\citenum{Eriksen2019a} from corresponding CASCI correlation energies, $\epsilon$, involving MO tuples, $\{p,q,r,\ldots\}$. To make calculations tractable, MBE-FCI implements both a screening and a purging protocol. In the former, MOs are screened away from the full expansion space at each order according to their maximum increment magnitude, which, in turn, leads to a reduced number of increment calculations at the orders to follow, while the latter purges away redundant increments at lower orders so as to avoid storing these in memory throughout the entirety of a calculation. Most recently, MBE-FCI has also been coupled to an optimized implementation of complete active space self-consistent field (CASSCF) theory where it replaces a standard CASCI solver~\cite{Greiner2024}.\\

The present study intends to address three major shortcomings of the current state of MBE-FCI for the calculation of electronic ground-state energies: $(i)$ the bias introduced by having to manually select upon reference active spaces, $(ii)$ the unfavorable scaling with the number of orbitals included in the complementary expansion space, and $(iii)$ the lack of rigour pertaining to the manner by which MOs are pruned from the expansion space. First, we are proposing a new adaptive algorithm capable of automatically detecting optimal active spaces based on the \textit{likeness} of individual CASCI wave functions and a parent reference, thus assuring that MBE-FCI calculations converge optimally onto realistic and correct electronic ground states. Second, we will discuss how to allow for a clustering of the orbitals of the expansion space in MBE-FCI, thus paving the way for highly accurate treatments of larger chemical systems. Third, we will outline a new screening protocol that operates by truncating the expansion space throughout an MBE subject to an {\textit{a priori}} error tolerance, enabling not only more rapid and robust convergences of expanded energies onto corresponding FCI solutions, but importantly also resulting error bars to be associated with these.\\

Through applications across the so-called {\texttt{FCI21}} benchmark set of small closed- and open-shell molecules by Chilkuri and Neese~\cite{neese_fci21_jctc_2021}, these new developments are tested for a variety of different MO bases; importantly, the correctness of the proposed error control is validated, as is the robustness of our new algorithm for detecting tailored reference active spaces. In fact, we further demonstrate how, in standard bases of either canonical or spatially localized MOs, active spaces are identified that align with well-established prior experience, and we show how the sizes of these may be controlled through the introduction of only a single, physically intuitive input parameter, thus warranting use cases beyond MBE-FCI, e.g., in multireference self-consistent field (MCSCF) theory~\cite{Greiner2024}. Finally, we demonstrate appropriate scaling of our orbital clustering and error control algorithms with system size by evaluating the accuracy of MBE-based coupled cluster (CC) calculations on a few larger systems.

\section{Theory}\label{theory_sect}

In here, we will discuss our new algorithm for the automatic detection of reference active spaces within MBE-FCI in Sect. \ref{ref_space_subsection}, while Sects. \ref{orbital_clustering} and \ref{error_screen_subsection} are concerned with the orbital clustering used throughout this work and our new dynamic screening protocol, respectively.

\subsection{Detection of Reference Active Spaces}\label{ref_space_subsection}

In the most general formulation of MBE-FCI theory~\cite{Eriksen2019a}, one opts for an empty (vacuum) reference space, in which case the complementary expansion space will be spanned by the complete set of MOs. At the leading order of such an MBE, all possible CASCI calculations that make reference to unique pairs of occupied and virtual orbitals are performed, while at subsequent orders, these pairs become augmented by additional single MOs of any occupation. However, in the case of open-shell systems, or whenever excited rather than ground states are targeted, one must retain certain MOs within a minimal reference space that need be included in all possible CASCI calculations to ensure convergence onto correct states.\\

Now, although such an algorithm will clearly remain unbiased towards any partitioning of a set of MOs, orbital-based MBEs will only converge satisfactorily whenever a given minimal reference (e.g., the HF determinant) stays dominant, not only in the FCI wave function, but also in all CASCI wave functions. For systems with electronic structures that instead demand particular emphasis on an active space of MOs, increment CASCI calculations that fail to make reference to any of the MOs of this active space will obviously be at risk of converging onto altogether erroneous electronic states. Even in the case of systems dominated by only a single determinant, some MOs might prove more important for the treatment of dynamic correlation than others. As is true for all active space methods, MBE-FCI will thus in general rely substantially on prior or educated knowledge of the chemistry at hand if treatments are to be successful. For that reason, we will now proceed to propose a simple algorithm for automatically detecting suitable reference spaces in the context of MBE-FCI. In Sect. \ref{results_sect}, we will further proceed to demonstrate how our selections of MOs will mirror conventional choices of active spaces for a number of well-established benchmark systems.\\

The new algorithm is intrinsically designed to be as unbiased as possible by initiating a given MBE from an empty reference space and leveraging the concept of {\textit{state fidelity}} known from quantum information theory (QIT) to measure the similarity of quantum states.\cite{Uhlmann1976, Alberti1983, Peres1984, Jozsa1994} This quantity has previously been applied to the field of quantum chemistry in the context of density matrix renormalization group (DMRG) theory.\cite{Boguslawski2011} For two pure quantum states, $I$ and $J$, the fidelity is given as the absolute square of the overlap between the two states:
\begin{align}
    F_{IJ} = \left|\braket{\Psi_I|\Psi_J}\right|^2 \ . \label{quant_fidel_eq}
\end{align}
During every increment calculation, the state found to produce the largest fidelity with the HF determinant is monitored, followed, and thus chosen as a reference. Most often, this will be the ground state within the active space constructed from a respective tuple of MOs. However, for multireference systems, the energetic ordering of states may swap upon successively adding orbitals to an active space, and the use of incorrect states in the recursive calculation of increments risks introducing major convergence issues throughout an MBE~\bibnote{We note how, when calculating the quantum fidelity of two CASCI states of which one active space is a subspace of the other, special care must be taken to account for the correct sign of the individual coefficients in Eq. \ref{quant_fidel_eq} in cases where the particle number in the two active spaces differs.}.\\

Whenever the fidelity of a chosen state drops below a given threshold (cf. the supporting information (SI) for specific details), this generally indicates a significant change with respect to the desired wave function. As a means to manage such a change, the algorithm will then include these particular MOs into the reference space and, consequently, restart the MBE based on this updated space instead. In turn, this switches the reference wave function from being comprised by simply the HF determinant to a new reference space CASCI wave function, against which state fidelities are then subsequently computed. In an automatic fashion, this procedure repeats itself until the MBE converges in a satisfactory fashion. In practice, the identification of possible reference space MOs takes place during the first few orders of the expansion only. As a result, the added computational overhead associated with the detection of an optimal reference active space becomes negligible in comparison to the cost of running MBE-FCI calculations through the higher expansion orders to follow.\\

For molecular systems characterized by dynamic correlation alone, the algorithm may decide to add no MOs to the reference space, whereas in the presence of static correlation, an empty reference space is most often extended by a selected set of MOs. In extreme cases, however, a given reference space may potentially grow excessively large. To avoid this issue, we have chosen not to implement a static threshold for the quantum fidelity, but instead an adaptive function that accounts for how large the reference space grows upon adding successive orbital tuples. Starting from an empty reference space, the first orbitals can be added at the second order. At this point, a sigmoid function is initialized with a given threshold value (usually $0.90$ to $0.95$), and the function then quickly drops as more and more orbitals are added. In practice, this prevents reference spaces from comprising more than 10 MOs or so. A dynamic threshold additionally also prevents reference spaces from being augmented by a large number of MOs at late orders due to an accumulation of dynamical correlation. As it happens, the addition of a large number of MOs will usually indicate that either a given orbital basis or the MBE-FCI method itself are unsuitable for the accurate treatment of a particular system (for instance, in the presence of strong correlation).

\subsection{Orbital Clustering}\label{orbital_clustering}

While truncated MBEs can accurately converge results towards the FCI limit for many systems not amenable to exact diagonalization, limits still exist in terms of the practical applicability of the MBE-FCI method whenever expansion spaces grow excessively large and the polynomial scaling of the number of increments becomes prohibitive. While efficient parallelization will aid somewhat, general restrictions will persist. Additionally, orbital-based MBE theory is not limited to the exact realm of FCI theory but can also be applied to approximate theories, e.g. truncated CC expansions, which is achieved by replacing the CASCI solver with a CAS solver for the quantum-chemical method of choice. Due to the polynomial scaling of conventional CC methods and the sheer number of increment calculations needed at higher orders, MBE-CC methods will find it difficult to compete with parent methods at the lower end of the CC hierarchy, such as CCSD or CCSD(T)~\cite{ccsd_paper_1_jcp_1982, original_ccsdpt_paper}. However, as standard reference results are easier to obtain at a given level of CC theory than with FCI, the formal convergence of general MBE-based treatments may be assessed using MBE-CC formulations.\\

To delay the onset of these problems intrinsic to all MBE-based methods, expansion objects based on clusters rather than single MOs may be used. It should be noted at this point how similar ideas have been pursued in the cluster MBE method by Abraham and Mayhall~\cite{Abraham2021}, which employs arbitrary clusters in conjunction with a cluster mean-field reference,\cite{JimenezHoyos2015} as well as in the iCASSCF method by Zimmerman and Rask~\cite{Zimmerman2019}, utilizing pairs of occupied and virtual valence MOs from a perfect-pairing reference.\cite{Hurley1953,Goddard1978,Langlois1990,Gerratt1997} Additionally, Dolg and co-workers have utilized occupied orbital clusters in conjunction with projected atomic orbital excitation spaces in incremental CCSD expansions.\cite{Friedrich2007} Orbital clustering furthermore bears some resemblance with the concept of $\pi$-pruning used in MBE-FCI to accelerate and ensure convergence onto the correct ground state for linear systems by assigning degenerate $\pi_x$ and $\pi_y$ orbitals to the same pair clusters~\cite{Eriksen2019,Eriksen2019a}. While orbital clustering does not alter the scaling of the MBE itself, the number of expansion objects is effectively reduced to avoid a combinatorial explosion at late orders. This reduction does, however, come at a price, given how the computational cost of individual increment calculations will increase, and choosing upon optimal orbital clusters thus involves a delicate balancing act between the number of increments and the scaling of the electronic-structure method used in the MBE.\\

When using orbital clusters as expansion objects, at least two natural definitions of the MBE order exist, either via the number of expansion objects considered per increment or in terms of the number of orbitals involved in individual CASCI calculations. In single-orbital MBEs, both of these definitions trivially coincide. Furthermore, both choices will produce identical MBEs whenever orbital clusters are homogeneous, that is, whenever the total number of orbitals is divisible by the desired cluster size. Unfortunately, this is generally not the case, and the need for heterogeneous clusters has made us opt for the second definition where the computational effort required across all increment calculations at a given MBE order is similar, thus working to improve load-balancing in parallel calculations.\\

Whenever screening protocols are introduced, it becomes beneficial for individual clusters to be designed such that those $n$-orbital contributions that appear at later orders in the MBE are insignificant. In the course of the present work, we have attempted to achieve this by exploiting single-orbital information at lower expansion orders to construct a set of orbital clusters that is as homogeneous as possible. The exact procedure for doing so is outlined in the SI. In our experience, whenever an MO basis is spanned by symmetry-adapted delocalized orbitals, such a procedure will place MOs of the same irreducible representation into clusters. With reference to the $\pi$-pruning discussed above for applications to linear molecules, the chosen clusters are thus on par with manual assignment whenever the provided cluster sizes permit for this. In the case of localized orbitals, the algorithm will generally work to cluster orbitals in close spatial proximity of one another, as is once again arguably to be desired.

\subsection{Error Control and Orbital Screening}\label{error_screen_subsection}

As mentioned in Sect. \ref{intro_sect}, for the  MBE-FCI method to be practically tractable, expansions must be enforced to terminate at some order significantly less than the size of the full expansion space. For this purpose, a screening protocol is implemented to truncate the orbital space, and MOs have hitherto been screened away according to their contributions at previous orders. At any given MBE order, only a fixed percentage of those MOs belonging to the expansion space that give rise to the numerically largest increments are retained at the orders to follow. An implication of this procedure is that the same relative reduction of the expansion space takes place for any given set of MOs, and any expansion will thus be forced to terminate regardless of whether or not it actually converges onto any target property.\\

This obvious weakness thus calls for a more physically motivated truncation scheme. At the same time, it would be desirable to encode some degree of error control into MBE-FCI, not only to base the screening of orbitals more stringently on the actual convergence behaviour of an MBE, but also to associate approximate uncertainties with final results. Beyond aspects concerned with the computational efficacy of MBE-FCI, a systematic procedure for screening out irrelevant increments at higher orders is motivated by the following two assumptions behind MBE convergences. First, the {\textit{mean absolute increment}} must decrease in value at a faster rate than the total number of increments at a given order increases. Second, the majority of increments will necessarily cancel due to similar magnitudes and {\textit{differing signs}} at later orders in the expansion. While the former of these two assumptions in itself warrants convergence, the latter will help to accelerate this behaviour.\\

\begin{figure}[htb]
    \includegraphics[width=0.8\textwidth]{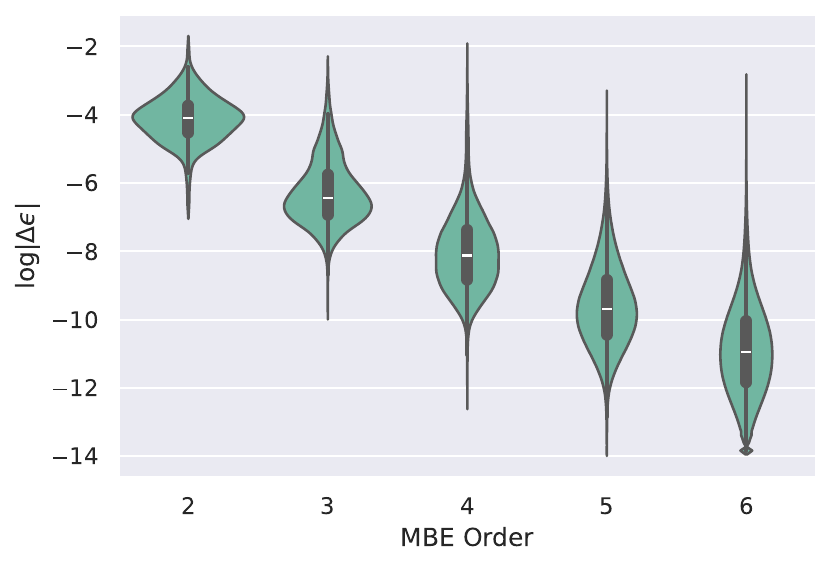}
    \caption{Distribution of log-transformed absolute increments ($\log|\Delta\epsilon|$) at different expansion orders ($k \leq 6$) for a single-orbital frozen-core MBE-FCI calculation of H$_2$O/cc-pVTZ based on an empty reference space. The symmetric kernel density estimate of the distributions is displayed in green, while the white lines indicate the median, the boxes the upper and lower quartiles, and the whiskers the data points within $150\%$ of the interquartile range.}
    \label{h2o_inc_convergence}
\end{figure}
In principle, the effects of both assumptions may be estimated by performing a random sampling of increments up through an MBE. Unfortunately, due to their recursive nature, such a strategy would necessitate prior knowledge of lower-order increments, cf. Eq. \ref{mbe_eq}. Thus, only increments at order $k+1$ may be trivially sampled at order $k$. Instead, we here opt for an extrapolation scheme based on an empirical relationship in combination with Monte-Carlo importance sampling~\cite{Goertzel1949, Kahn1951}, a strategy which collectively allows for truncations of MBEs to within a given error bound. As can be seen from the results for H$_2$O in Fig. \ref{h2o_inc_convergence}, absolute increments, $|\Delta\epsilon|$, decrease exponentially with the expansion order. For a given order, increments are log-normally distributed with mean values decreasing linearly with expansion order, thus permitting extrapolation. As the majority of increments approaches the convergence criterion of the individual CASCI calculations ($10^{-10}\ E_\mathrm{h}$ by default), individual increments still appear to adhere to this distribution. However, near the numerical threshold for double-precision floating point arithmetics, distributions are naturally observed to deteriorate slightly. Analogous results for NH$_3$ and CH$_4$ are presented in Figs. S6 and S7 of the SI.\\

When instead using clusters rather than single MOs as expansion objects, the relatively simple relationship visible from Fig. \ref{h2o_inc_convergence} becomes more complicated, given how different combinations of clusters can group and contribute at a particular order in an MBE. For instance, the incremental energy contribution from correlating four MOs amongst each other can result from the union of two pair clusters, from a pair alongside two single MOs, and from four single MOs. With our definition of the MBE order (cf. Sect. \ref{orbital_clustering}), the magnitude of an increment is thus not necessarily correlated with the order itself as there will exist correlations among the orbitals of clusters that have not been accounted for earlier on in the expansion. Again, for the illustrative example of an increment associated with any four MOs (order $k=4$), when two MO pairs give rise to this increment, all 2-orbital correlations between the clusters will not have been described at earlier orders. Here, the number of clusters, $N_\mathrm{cluster}$, will be 2, and there will be a total of 4 of such contributions, $N_\mathrm{contrib}$, determined as the product of the individual cluster sizes. Likewise, when an increment at $k=4$ results from one pair and two single MOs, $N_\mathrm{cluster} = 3$ and $N_\mathrm{contrib} = 2$, while $N_\mathrm{cluster} = 4$ and $N_\mathrm{contrib} = 1$ when an increment accounts for the residual correlation among any four single MOs.\\

\begin{figure}[htb]
    \includegraphics[width=0.8\textwidth]{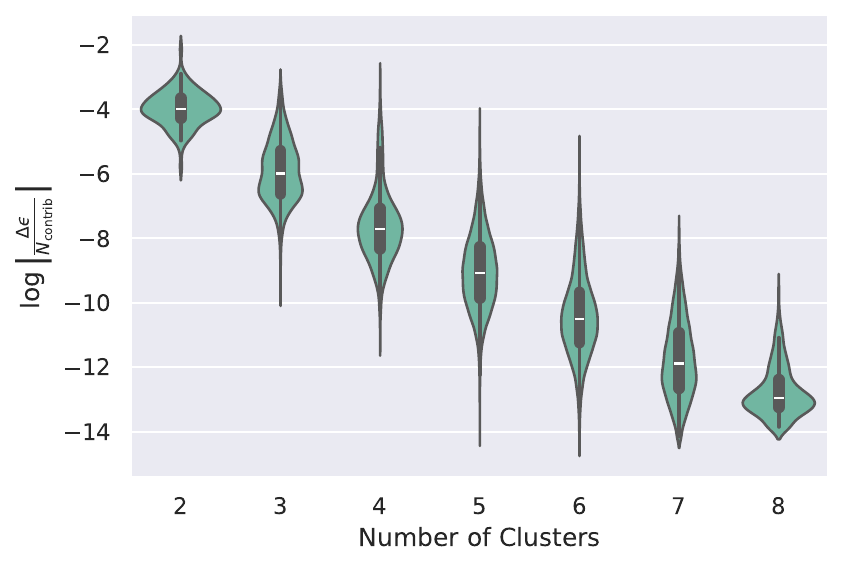}
    \caption{Distribution of log-transformed normalized absolute increments ($\log|\Delta\epsilon / N_\mathrm{contrib}|$) at different MBE orders ($k \leq 8$) for the same H$_2$O/cc-pVTZ example as in Fig. \ref{h2o_inc_convergence}. Orbital clusters comprising between 1 and 4 MOs have been selected upon at random.}
    \label{h2o_cluster_inc_convergence}
\end{figure}

On the basis of the discussion above, we have chosen to generally describe the magnitude of an increment through the following empirical relationship:
\begin{align}
|\Delta\epsilon| = aN_\mathrm{contrib} \cdot  \exp{(bN_\mathrm{cluster})} + \bar{\varepsilon} \ . \label{regression_equation}
\end{align}
In Eq. \ref{regression_equation}, $\bar{\varepsilon}$ describes the uniform error inherently present in all increments due to differing occupation and orbital interaction. The increment magnitudes in Eq. \ref{regression_equation} can be normalized by dividing by the number of contributions, the distributions of which are displayed in Fig. \ref{h2o_cluster_inc_convergence}.\\

As for the single-orbital case in Fig. \ref{h2o_inc_convergence}, the magnitudes of the increments in Fig. \ref{h2o_cluster_inc_convergence} are distributed log-normally with the mean decreasing exponentially upon an increase in the number of clusters (cf. also Figs. S8 and S9 of the SI). Again, normalized increments may hit a practical limit resulting in censored normal distributions, as is visible in the results for increments involving 7 or more clusters in Fig. \ref{h2o_cluster_inc_convergence}, producing distributions that are slightly skewed and shifted to higher energies. The SI provides a more detailed discussion on the effects of finite convergence criteria in relation to this matter. While the apparent scaling of mean increments is evident from Fig. \ref{h2o_cluster_inc_convergence}, the variance of their magnitudes for a given number of clusters will remain substantial, rendering the predictive powers of an empirical model based directly on Eq. \ref{regression_equation} for screening and error estimation purposes only marginal at best.\\

\begin{figure}[htbp]
    \includegraphics[width=0.95\textwidth]{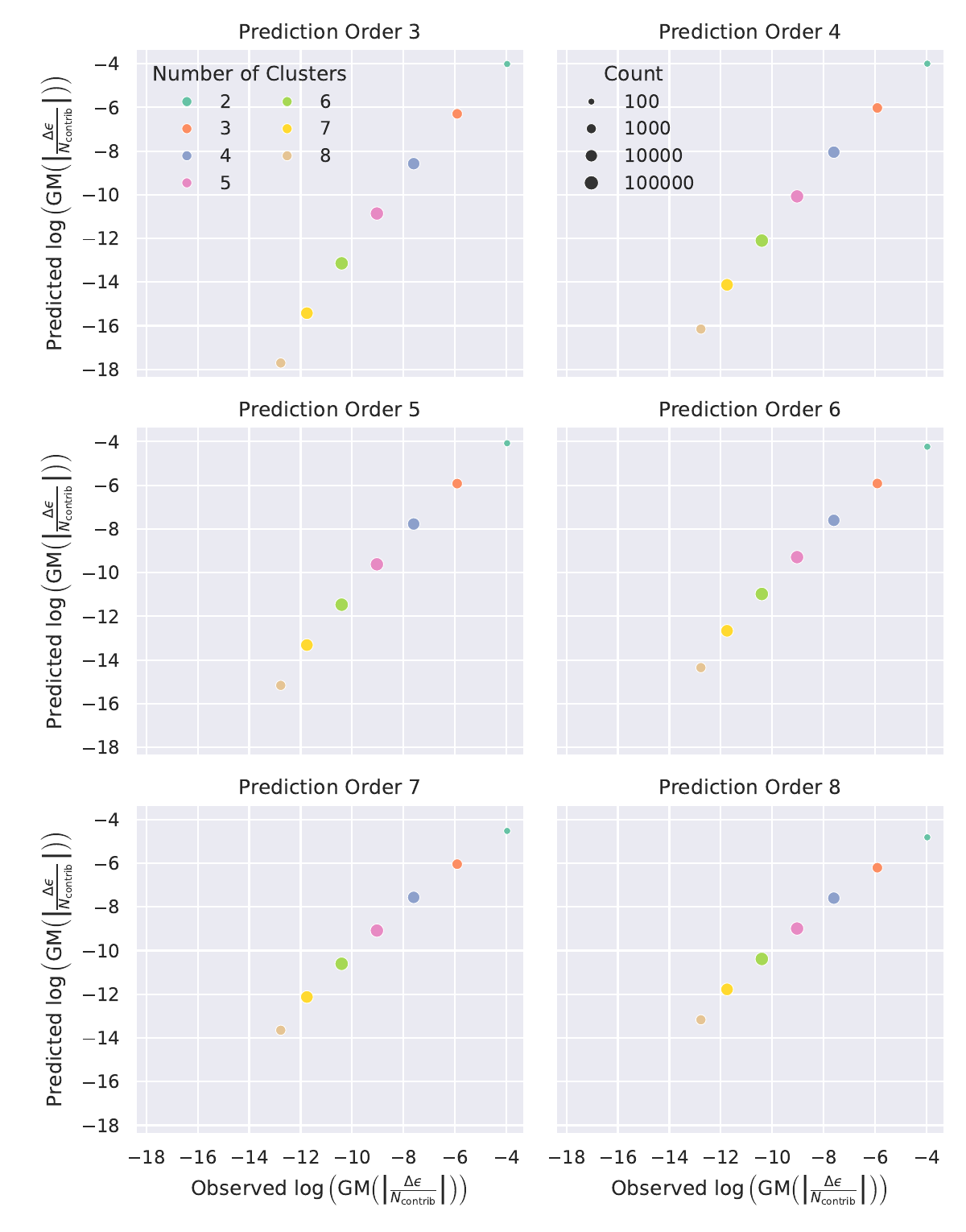}
    \caption{Predicted vs. observed logs of the geometric mean (GM) of normalized increment magnitudes at different prediction orders for the H$_2$O/cc-pVTZ example in Figs. \ref{h2o_inc_convergence} and \ref{h2o_cluster_inc_convergence}.}
    \label{h2o_fit_quality}
\end{figure}
For this reason, we have instead designed a machine model based on a weighted linear regression of the geometric mean (GM) of the normalized increment magnitude in the log-transformed scale, cf. Fig. \ref{h2o_fit_quality} (alongside Figs. S10 and S11 of the SI). Predictions at early MBE orders will generally tend to underestimate observed increment magnitudes due to deviations from linearity caused by finite convergence criteria and limited floating-point precision. Nevertheless, whenever the model is used for screening purposes, prediction intervals are added to the predicted GMs to ensure that actual means are rarely underestimated in practice. The SI provides more details on the exact use of the proposed model.\\

For log-normal distributions, the arithmetic mean (AM) can be determined from the corresponding AM and variance of normal distributions in the log-transformed scale, given how the AM in the log-transformed scale is equivalent to the logarithm of the GM, and variances can trivially be used from previous data points. By adding predicted AMs of increment magnitudes for all combinations of $N_\mathrm{contrib}$ and $N_\mathrm{cluster}$ produced by a particular orbital cluster at a given order $k$, these AMs can be related to the {\textit{potential}} total absolute contribution of said orbital cluster at order $k$, 
\begin{align}
    |\mathcal{E}^{(k)}| = \sum_{\Delta\epsilon} N_\mathrm{contrib}(\Delta\epsilon)\cdot\mathrm{AM}\left(\left|\frac{\Delta\epsilon}{N_\mathrm{contrib}}\right|\right)_{N_\mathrm{cluster}(\Delta\epsilon)} \ . \label{pot_contrib_eq}
\end{align}
Potential contributions from Eq. \ref{pot_contrib_eq} will equal {\textit{actual}} total absolute contributions of an orbital cluster in question, $E^{(k)}$, whenever no increments cancel, in which case accurate fits of GMs and perfect adherence to the log-normal distribution would permit for exact error bars to be added to our MBE-based results. In practice, however, potential absolute orbital contributions will far exceed actual ones ($0 < F_{k} < 1$):
\begin{align}
|E^{(k)}|=F_k\cdot|\mathcal{E}^{(k)}| \ . \label{tot_contrib}
\end{align}

Unfortunately, the degree to which cancellation occurs among increments at any given order is inherently more difficult to predict than the values of $\mathcal{E}^{(k)}$ through Eqs. \ref{regression_equation} and \ref{pot_contrib_eq}. A fair assumption is that $F_k = 1$ at the start of an MBE treatment, given how all individual increments are negative, while $F_k \to 0$ at higher orders whenever distributions of increments become symmetric and centered around zero value. As magnitudes of individual increments decrease, the potential for signs ($\pm$) to be flipped when additional $n$-orbital correlations are considered increases until signs eventually become equally distributed.\\

To obtain a conservative, overall estimation of increments, we thus choose to sample predicted distributions of mean values drawn from Eq. \ref{regression_equation} alongside approximate variances from distributions of increments at earlier orders by means of importance sampling. The required probabilities are acquired through kernel density estimation (KDE).\cite{Rosenblatt1956,Parzen1962} Increments are hence sampled according to their magnitude, while the signs of these are summed to determine an approximation to Eq. \ref{tot_contrib}. The sampling is repeated until convergence of the quantile of interest is met, and we again refer to the SI for more algorithmic details.\\

In summary, our new screening algorithm requires only a single input argument, namely, a maximum target error, $\gamma$. Through orbital-based fits of GMs, the algorithm is designed to decide whether an expansion is to be truncated or not at a given order only if final results may still be produced to within a predefined error tolerance. As a direct implication, more screening will take place for MBEs spanned in sets of MOs more fit for purpose than others (e.g., spatially localized MOs). Expansions will still terminate whenever no further increments can be constructed from the MOs of the expansion space, and we may now also associate final results with error bars as accumulated predictions from all screened orbitals.

\section{Computational Details}\label{comp_sect}

All MBE-based FCI and CC calculations have been carried out using the embarrassingly parallel {\texttt{PyMBE}} code~\cite{Eriksen2017, Eriksen2018, Eriksen2019, Eriksen2019a, eriksen2020, eriksen2021, pymbe}, using either the {\texttt{fci}} and {\texttt{cc}} modules in {\texttt{PySCF}}~\cite{Sun2017, Sun2020} or the {\texttt{ecc}} and {\texttt{ncc}} modules in {\texttt{CFOUR}}~\cite{Matthews2020, cfour} as electronic-structure kernels. KDEs from {\texttt{SciPy}} have been used for importance sampling purposes, with Scott's rule for bandwidth estimation~\cite{Virtanen2020,Scott1992}.\\

The abilities of our improved MBE-FCI method to produce results to within given target error bounds have been assessed on the {\texttt{FCI21}} benchmark set using the standard cc-pVDZ and SV basis sets~\cite{Dunning1989,Weigend2005}. A couple of changes to the {\texttt{FCI21}} set were made to adapt it to the purposes of the present study in that the calculation on \ce{H2}, the all-electron calculations on \ce{Li2} and \ce{Be2}, as well as the frozen-core calculation on \ce{HF} were removed. \ce{H2} obviously converges extremely quickly, while \ce{Li2} and \ce{Be2} are already represented by the respective frozen-core calculations that correspond with the calculations on the other homonuclear diatomics in the set. Likewise, \ce{HF} is already represented by an all-electron calculation which is in line with the calculations of the other diatomic hydrides. For \ce{NH} and \ce{B2}, the triplet states and the respective equilibrium bond lengths were used as the ground states. The corresponding parameters are summarized in Table \ref{fci21_parameters_table}. Throughout our study, converged MBE energies are compared to corresponding reference FCI and CC results calculated using {\texttt{PySCF}}.\\
\addtolength{\tabcolsep}{-0.2pt}
\begin{table}[htbp]
    \centering
    \begin{tabular}{cccccccc} 
        \hline
        Molecule & Bond Length         & Angle            & Dihedral Angle & Ground & Correlated & Basis \\
                 & (in \textup{\AA})   & (in degree)      & (in degree)    & State  & Orbitals   &       \\
        \hline
        \ce{LiH} & \num{1.6136} & ---         & ---                      & $^1\Sigma^+$            & all-electron & cc-pVDZ \\
        \ce{BeH} & \num{1.3570} & ---         & ---                      & $^2\Sigma^+$            & all-electron & cc-pVDZ \\
        \ce{BH}  & \num{1.2551} & ---         & ---                      & $^1\Sigma^+$            & all-electron & cc-pVDZ \\
        \ce{CH}  & \num{1.1424} & ---         & ---                      & $^2\Pi$                 & all-electron & cc-pVDZ \\
        \ce{NH}  & \num{1.0557} & ---         & ---                      & $^3\Sigma^-$            & all-electron & cc-pVDZ \\
        \ce{OH}  & \num{0.9796} & ---         & ---                      & $^2\Pi$                 & all-electron & cc-pVDZ \\
        \ce{HF}  & \num{0.9200} & ---         & ---                      & $^1\Sigma^+$            & all-electron & cc-pVDZ \\
        \ce{Li2} & \num{2.7139} & ---         & ---                      & $^1\Sigma_\mathrm{g}^+$ & frozen-core & cc-pVDZ \\
        \ce{Be2} & \num{4.4269} & ---         & ---                      & $^1\Sigma_\mathrm{g}^+$ & frozen-core & cc-pVDZ \\
        \ce{B2}  & \num{1.6240} & ---         & ---                      & $^3\Sigma_\mathrm{g}^-$ & frozen-core & cc-pVDZ \\
        \ce{C2}  & \num{1.2728} & ---         & ---                      & $^1\Sigma_\mathrm{g}^+$ & frozen-core & cc-pVDZ \\
        \ce{N2}  & \num{1.1368} & ---         & ---                      & $^1\Sigma_\mathrm{g}^+$ & frozen-core & SV      \\
        \ce{O2}  & \num{1.2786} & ---         & ---                      & $^3\Sigma_\mathrm{g}^-$ & frozen-core & SV      \\
        \ce{F2}  & \num{1.4186} & ---         & ---                      & $^1\Sigma_\mathrm{g}^+$ & frozen-core & SV      \\
        \ce{CH4} & \num{1.1015} & \num{109.5} & \num{120.0}, \num{240.0} & $^1A_1$                 & frozen-core & SV      \\
        \ce{NH3} & \num{1.0277} & \num{103.5} & \num{107.7}              & $^1A_1$                 & frozen-core & cc-pVDZ \\
        \ce{H2O} & \num{0.9668} & \num{101.9} & ---                      & $^1A_1$                 & frozen-core & cc-pVDZ \\
        \hline
    \end{tabular}
    \caption{Systems from the {\texttt{FCI21}} benchmark set used in the present work~\cite{neese_fci21_jctc_2021}.}
    \label{fci21_parameters_table}
\end{table}
\addtolength{\tabcolsep}{0.2pt}

For all systems in Table \ref{fci21_parameters_table}, optimal reference spaces have been automatically identified based on a single-orbital MBE-FCI calculation truncated at order $k = 5$. Expansion spaces were next divided into clusters of at most two MOs, and results have been compared for four typical choices of MOs: $(i)$ canonical orbitals, $(ii)$ CCSD natural orbitals (CCSD NOs), as well as $(iii)$ Pipek-Mezey~\cite{Pipek1989} (PM) and $(iv)$ Foster-Boys~\cite{Foster1960} (FB) localized molecular orbitals (LMOs). The SI reports further results with an additional six types of MOs, for instance, split-type and CASSCF bases. Restricted (open-shell) HF states of correct multiplicity and symmetry were used throughout with the exception of \ce{OH}, for which the broken-symmetry $B_2$ solution was used instead. Similarly, in the case of the calculations on both \ce{CH} and \ce{OH}, the CCSD solutions used to compute NOs were also of $B_2$ symmetry, as were the corresponding CASSCF solutions for these two systems (cf. the results in the SI)~\bibnote{As the {\texttt{PySCF}} FCI solver is restricted to Abelian and linear point groups, the $A'$ state was targeted during the CASSCF optimization for the \ce{NH3} and \ce{CH4} systems.}.\\

Our algorithm for detecting reference spaces (Sect. \ref{ref_space_subsection}) has been tested on three systems with some degree of static correlation: ozone, nickel acetylene, and the phenoxy radical. The sequence in which MOs were successively added to reference active spaces was compared to that governed by single-orbital entropies and mutual information used in other active space selection schemes~\cite{Stein2016, Stein2019}. These quantities were calculated from DMRG wave functions generated using the {\texttt{BLOCK2}} code~\cite{Zhai2023}, using a small bond dimension of 500 renormalized block states. The geometries for these systems were taken from the work of Keller {\textit{et al.}}~\cite{Keller2015}, and the active space selection was applied to the full valence spaces in the cc-pVTZ basis set.~\cite{Dunning1989}\\

To demonstrate the effectiveness and applicability of our clustering (Sect. \ref{orbital_clustering}) as well as screening and error estimation protocols (Sect. \ref{error_screen_subsection}), the algorithms are finally tested on a small group of larger systems in a cc-pVTZ basis set on the basis of CCSD geometries: \ce{C2H4}, \ce{CH3OH}, \ce{H2CO}, \ce{H2O2}, \ce{HCN}, and \ce{N2H4}. Since conventional FCI is intractable for problems of this size, errors of the MBE-CCSD method are instead compared to conventional CCSD.

\section{Results}\label{results_sect}

\begin{figure}[htbp]
    \includegraphics[width=\textwidth]{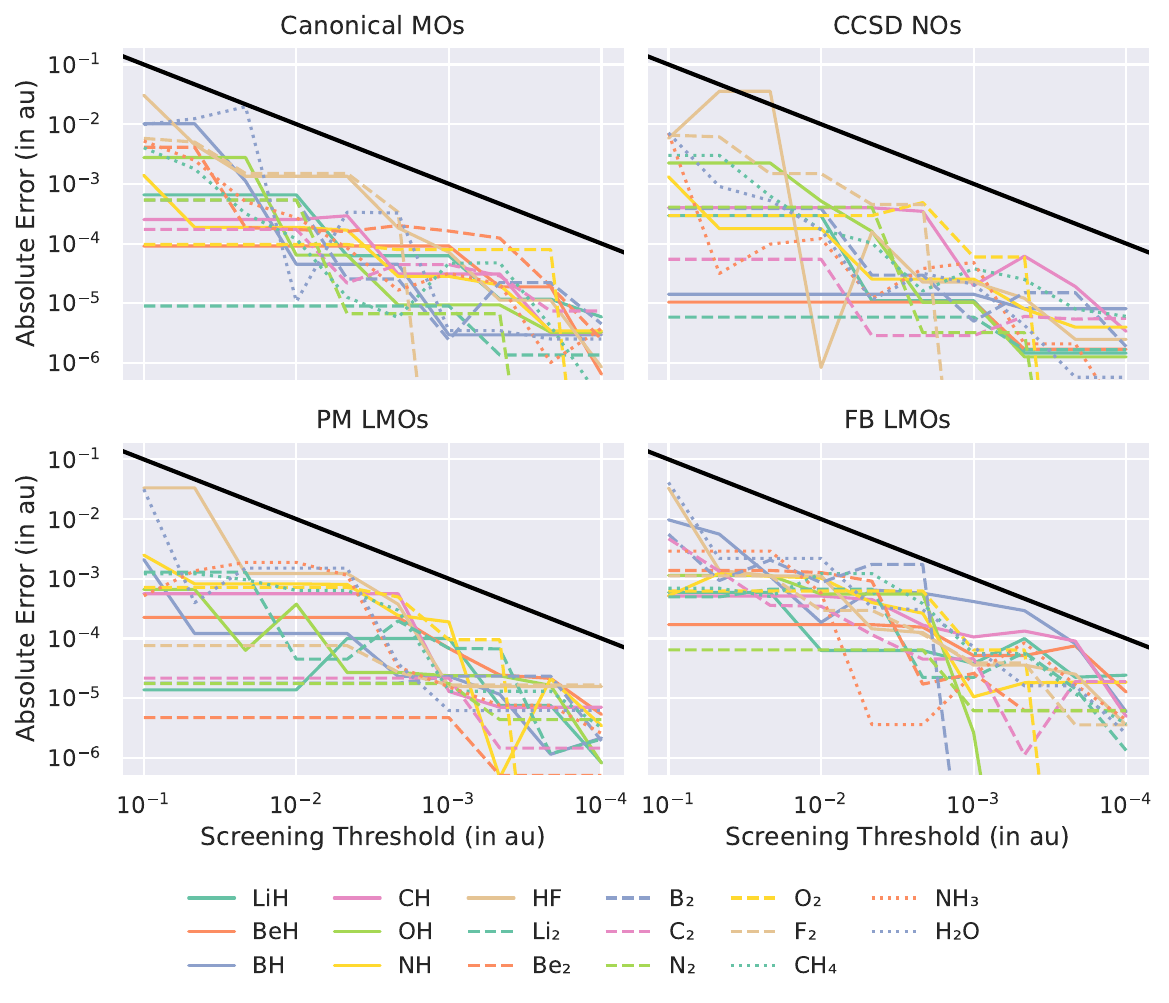}
    \caption{Absolute errors across the {\texttt{FCI21}} set for different screening thresholds and MO bases. Black lines indicate perfect correspondence between actual and predefined errors.}
    \label{benchmark_error}
\end{figure}

On account of the fact that our screening procedure is based on a predefined error threshold, the ability of the method to produce results to within this error bound is crucial and obviously needs to be numerically verified. In Fig. \ref{benchmark_error}, convergence properties are assessed on the modified {\texttt{FCI21}} benchmark set (cf. Table \ref{fci21_parameters_table}) using the four MO bases discussed in Sect. \ref{comp_sect}.\\

The results in Fig. \ref{benchmark_error} consistently demonstrate how our algorithm is capable of yielding results to within desired error bounds across all systems covered in this benchmark. As a general trend, errors decrease as the screening threshold is tightened towards $0.1 \ mE_\mathrm{h}$. Final errors are found to only rarely exceed the bounds, caused by a lack of statistical significance due to a limited number of increments for these small systems expressed in modest basis sets.\\

\begin{figure}[htb]
    \includegraphics[width=\textwidth]{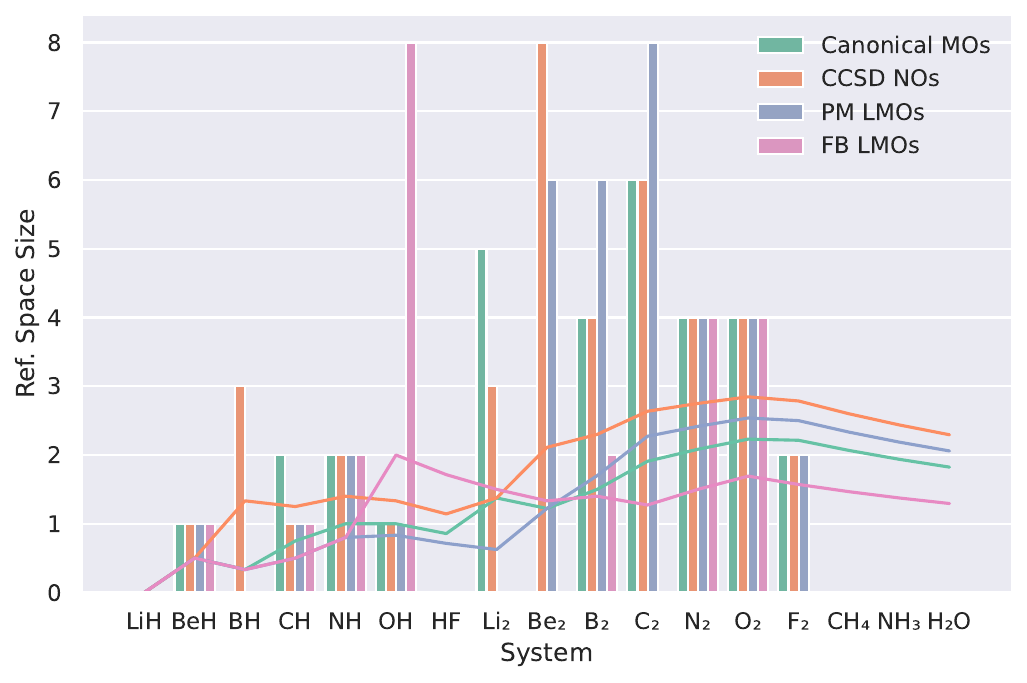}
    \caption{Sizes of reference active spaces expressed in different orbital bases across the {\texttt{FCI21}} set. The bars indicate sizes of individual reference spaces for the various systems, while the horizontal lines depict the corresponding cumulative average size of the reference spaces.}
    \label{benchmark_ref_spaces}
\end{figure}
For the calculations in Fig. \ref{benchmark_error}, a maximum quantum fidelity threshold of 0.95 was used (cf. Eq. \ref{quant_fidel_eq}), and reference spaces were selected from an initial single-orbital MBE-FCI calculation with orders $k \leq 5$. The sizes of these are displayed in Fig. \ref{benchmark_ref_spaces}. The diatomic hydrides are found to predominantly require no reference spaces beyond what is possibly dictated by their open-shell nature. However, for the OH system in the FB basis, the algorithm rules the inclusion of a total of 8 reference MOs. Homonuclear diatomics in the benchmark set tend to require orbitals to be added to the reference spaces as their ground-state wave functions often have non-vanishing contributions from more than a single determinant, e.g., \ce{Be2}, \ce{B2}, and \ce{C2}, which all exhibit some multireference character. Finally, in line with expectation, no reference spaces tend to be required for MBE-FCI calculations on systems dominated by dynamic correlation (\ce{LiH}, \ce{HF}, \ce{CH4}, \ce{NH3}, and \ce{H2O}), regardless of the choice of MO basis.\\

\begin{figure}[htb]
    \includegraphics[width=\textwidth]{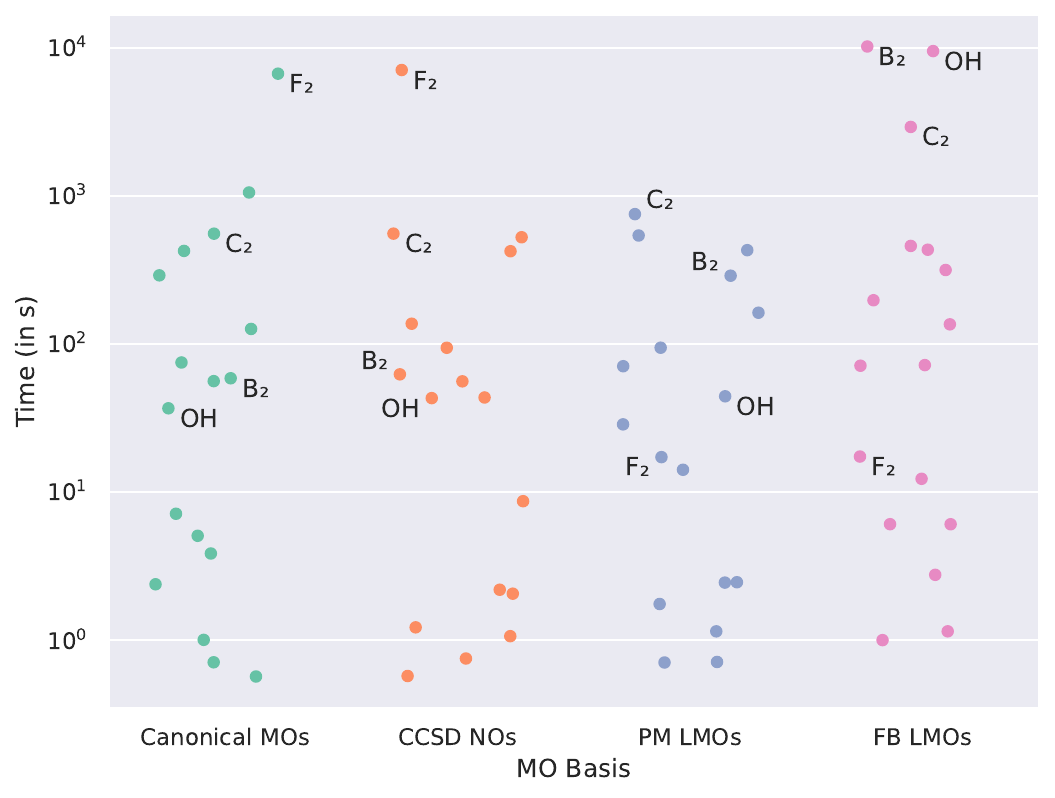}
    \caption{MBE-FCI timings on the {\texttt{FCI21}} set for different choices of MO basis.}
    \label{benchmark_timings}
\end{figure}

The effects on both MBE convergence and time-to-solution from using different MO bases and the corresponding automatically identified reference spaces are somewhat unclear from the sizes of these spaces alone. Generally speaking, the addition of MOs to a reference space will work to accelerate convergence throughout an MBE by decreasing the amount of increments that need be evaluated, but this is obviously accompanied by a simultaneous increase in the computational cost associated with each individual CASCI calculation. To investigate things further, Fig. \ref{benchmark_timings} compares wall times involved in converging MBE-FCI calculations across the {\texttt{FCI21}} set to within an error bound of $1 \ mE_\mathrm{h}$ for the same MO bases as in Fig. \ref{benchmark_ref_spaces}. Timings for a larger variety of orbital combinations are presented in the SI~\bibnote{Computer architecture: 4 Intel Xeon CPUs E5-4620 on a single node (64 threads, 32 cores @ 2.4 GHz, 7.80 GB/thread)}.\\

In addition to effects related to reference spaces, and the implications these have on the number of increments as well as the cost of their evaluations, certain choices of MO bases, particularly those localized in space, will work to accelerate convergence throughout an MBE in a given expansion space by allowing for individual orbital clusters to describe more exclusive correlations. Likewise, some MO bases will allow for folding more correlation into a given reference space than others, although the use of very delocalized MOs in the construction of reference spaces may risk reducing the potential to localize the remaining MOs that span the complementary expansion spaces. Finally, we note how one should generally be cautious about drawing too many conclusions from comparisons on small, compact systems like those of the {\texttt{FCI21}} set, as different rotations of a given MO basis may not differ too much from one another. Comparisons are arguably better done for larger systems, for which the effect of spatial locality will be more pronounced, or systems dominated to a greater extent by static correlation, for which optimized (CASSCF) reference spaces will be favoured. Such systems will, however, in general not permit for direct comparisons to conventional (exact) FCI.\\

Smaller quantum fidelities for a larger number of MOs will produce larger reference spaces, and---as is shown in Fig. \ref{benchmark_ref_spaces}---variations in the composition of these exist for the four MO bases in question across the {\texttt{FCI21}} set, despite most species being diatomics. In reference to the detailed results in Fig. \ref{benchmark_ref_spaces}, timings for four specific species have been annotated in Fig. \ref{benchmark_timings}, namely, \ce{F2}, \ce{B2}, \ce{C2}, and \ce{OH}, with individual results for all species reported in the SI. In the case of \ce{F2}, all MO bases but that of FB LMOs yield a reference space of 2 orbitals. Be that as it may, MBE convergence is accelerated in a basis of PM/FB LMOs by almost three orders of magnitude, revealing how the time required to converge a given MBE relies just as intricately on the ability to compress correlations into very compact MO clusters as it does on the size of a particular reference space. That being said, the use of a reference space may affect wall times both positively and negatively. For the calculations on the \ce{OH} radical, the relatively large space of 8 MOs detected in the basis of FB LMOs substantially increases the total time-to-solution, while the small (2 MOs) and empty reference spaces for \ce{B2} and \ce{C2}, respectively, similarly suppress the convergence of MBE-FCI in comparison to the three other MO bases. Across all MO bases tested here (cf. also the SI), PM LMOs arguably emerge as the optimal choice, also given how inefficiently performance in the more delocalized canonical MOs and CCSD NOs will generalize when moving towards larger systems.\\

Now, how do reference spaces selected in MBE-FCI (on the basis of similarities between different quantum states) compare to active spaces central to traditional multireference methods? The identification of active spaces for CASSCF calculations often requires user input based on chemical intuition. This is known to impede the black-box character of such methods, particularly so whenever the orbitals of an initial MO basis cannot be reasonably sorted by energy or occupation. For this reason, a variety of automated schemes for optimal active space selection have been proposed over the years~\cite{Jensen1988,Bofill1989,Keller2015,Bensberg2023,Sayfutyarova2017,Sayfutyarova2019,Khedkar2019}, which all tend to leverage information gathered from more affordable calculations on the full system. One such example is {\texttt{autoCAS}}~\cite{Stein2016, Stein2019}, in which active MOs are identified based on single orbital entropies and their mutual information derived from DMRG calculations. On that note, Stemmle and Paulus have previously studied how MBE energy increments relate to measures of entanglement~\cite{Stemmle2019}.\\

\begin{figure}[htb!]
    \includegraphics[width=\textwidth]{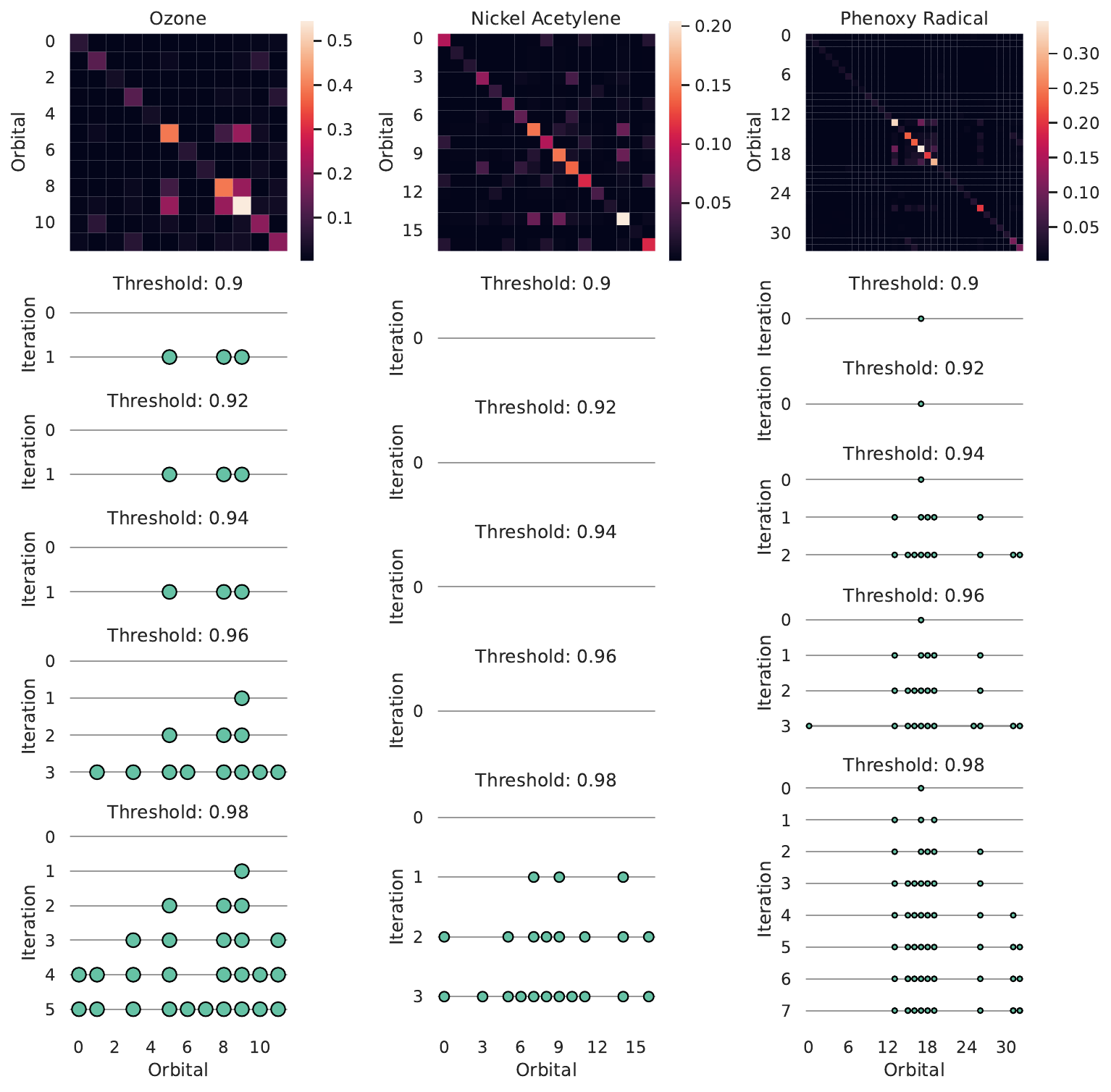}
    \caption{Comparison of quantum entanglement diagrams from DMRG calculations to iterative MBE-based addition of active space orbitals based on quantum fidelities for ozone, nickel acetylene, and the phenoxy radical in an PM LMO basis. Top: Diagonal elements describe single-orbital entropy, while off-diagonal elements describe mutual information. Bottom: Green dots indicate orbitals included in the active space at the end of an MBE iteration.}
    \label{entanglement_local}
\end{figure}
In Fig. \ref{entanglement_local}, we compare the iterations (number of MBE-FCI restarts) at which orbitals are added to different reference active spaces in our selection algorithm. At iteration 0, calculations are started with an empty reference space except for the phenoxy radical, for which singly occupied orbitals are always included in the reference space to ensure that the correct spin state can be targeted for every possible combination of orbitals. Again, higher quantum fidelity thresholds will cause more orbitals to be added to the active space. From the results in Fig. \ref{entanglement_local}, significant correlations are observed between the quantum fidelity of states resulting from different MO combinations and the orbital entropies and mutual information of these same MOs derived from DMRG. Orbitals with high entropies are typically added to our reference spaces earlier on and at lower thresholds, while orbitals forming a pair with high mutual information tend to get added simultaneously, as can be seen, for instance, for ozone and nickel acetylene at iteration 1 (notice the different scales in the three heatmaps).\\

In some cases at lower thresholds, it may happen that seemingly unimportant orbitals are added alongside more important ones as they manage to push the quantum fidelity over the threshold. At higher thresholds, however, the important orbitals of such MO tuples get added earlier on, while the unimportant orbitals are not added at all, cf. also the corresponding results in the SI based on canonical MOs. As discussed in Sect. \ref{ref_space_subsection}, we are largely circumventing this issue altogether by not using static quantum fidelities, as in Fig. \ref{entanglement_local}, but rather a dynamic protocol for reducing thresholds upon an increase of the reference space.\\

\begin{figure}[htbp]
    \includegraphics[width=\textwidth]{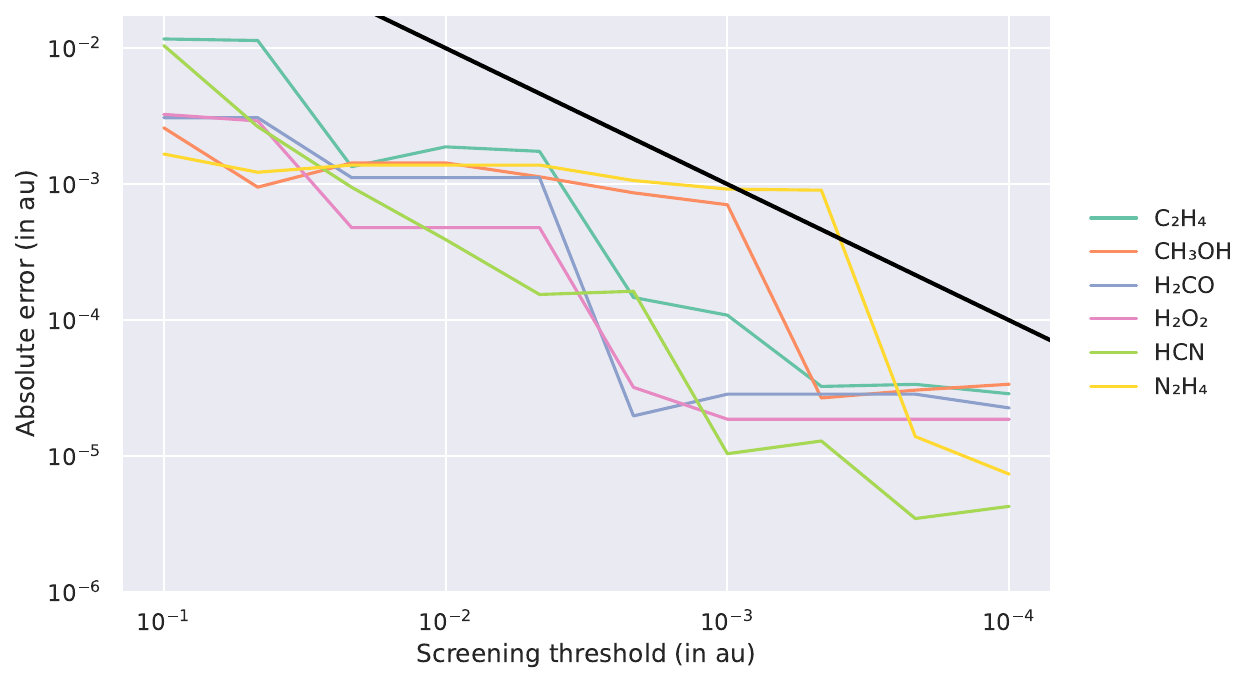}
    \caption{Absolute errors of MBE-CCSD/cc-pVTZ calculations on \ce{C2H4}, \ce{CH3OH}, \ce{H2CO}, \ce{H2O2}, \ce{HCN}, and \ce{N2H4} for different screening thresholds using bases of PM LMOs.}
    \label{mbe-ccsd_error_stats}
\end{figure}
Finally, we are now in a position to assess how the performance of our revamped method will scale to larger systems. As mentioned in Sect. \ref{comp_sect}, we will do so by applying MBE-CCSD to a group of chemical systems for which FCI is intractable and instead compare final energies to conventional CCSD. Errors of all these expansions for different screening thresholds are plotted in Fig. \ref{mbe-ccsd_error_stats}, on par with Fig. \ref{benchmark_error}. In each and every calculation, we use empty reference spaces and automatically generated orbital clusters of size 8 based on PM LMOs.\\

For the systems in Fig. \ref{mbe-ccsd_error_stats}, MBE-CCSD is consistently able to produce energies close to the predetermined error bound, thus corroborating the results on the smaller systems of the {\texttt{FCI21}} set. Given how our screening procedure requires a certain amount of statistical information to set in, expansions will often produce energies well below the threshold before any screening of orbitals takes place. In this regime, $F_k = 1$ in Eq. \ref{tot_contrib} and the upper error bound is correctly estimated whenever predicted mean absolute increments are reasonably approximated. At later orders, actual errors fall much closer to the predicted errors because increment magnitudes will typically not observe exact log-normal distributions and tails in these will start to affect arithmetic mean values, cf. the discussion in Sect. \ref{error_screen_subsection}.\\

For lower error bounds, many increments at late MBE orders will fall close to or below the convergence criterion used in individual increment calculations (cf. the SI). Such artefacts are bound to introduce some degree of noise into both our screening and error estimations. This is inherently less of a problem for MBE-FCI than it is for MBE-CC due to different convergence patterns exhibited by the individual increment calculations. In both cases, however, increment magnitudes will still be bounded from below by the limitations of finite floating-point arithmetics. For this reason, there will be natural restrictions on the accuracy that can be achieved through MBE-based methods, which can only be alleviated by either increasing cluster sizes or enabling floating point numbers of higher precision. 

\section{Discussion and Conclusions}

We have presented several key improvements to the MBE-FCI method that collectively work to enhance the ability to treat electronic structures in significantly more robust and black-box manners. The proposed automatic detection of reference active spaces offers an unbiased treatment of chemical systems that involve both dynamic and static electron correlation while ensuring optimal convergence of MBEs toward a given target property of choice. We have demonstrated how such reference spaces agree well with active spaces identified by means of quantum entanglement measures. In the treatment of electronic ground states, this is accomplished by initially targeting the state that shows the maximal quantum fidelity with the HF Slater determinant. However, the current procedure may also be extended to the treatment of excited states by starting from a reference space able to describe a specific determinant which dominates a CIS wave function and following this target state instead.\\

While the orbital clustering algorithm is capable of extending the applicability of MBE-based FCI and CC methods to larger systems and basis sets, the choice of these orbital clusters may be improved further by explicitly taking variable cluster sizes and occupations into account. Additionally, MBE-FCI may be extended to allow for overlapping clusters which would enable additional intra-cluster contributions at early expansion orders, akin to similar ideas previously applied to fragment-based MBEs of condensed-matter systems~\cite{Richard2012}. Finally, we are currently working on combining the present developments with the efficient use of (non-)Abelian point-group symmetry in MBE-FCI by means of tailored MOs~\cite{Greiner2023}.\\

The current screening and error estimation protocol constructs log-normal distributions from the geometric mean of all increments, regardless of whether these might lie below the convergence criterion of individual CASCI calculations. Instead, such increments could potentially be treated as censored, in which case resulting distributions could be determined through maximum likelihood estimation~\cite{Cox2018}. This explicit treatment could overcome precision loss in the error estimation whenever the geometric mean increment magnitude approaches the convergence criteria of the CASCI calculations. We foresee that this would only become an issue of practical relevance in the context of thermochemical calculations with error bounds beyond $0.1 \ mE_\mathrm{h}$. Either way, for the outlined error estimation to provide strict upper bounds will stay contingent on the observed, empirical relationships discussed herein.

\section*{Acknowledgments}

This paper is dedicated to Professor Trygve Helgaker on the occasion of his 70th birthday in 2023. J{\"u}rgen Gauss wishes to thank Trygve for almost 30 years of friendship, his great hospitality during many visits to Oslo (including a sabbatical and a stay at the {\textit{Centre for Advanced Study}} at the {\textit{Norwegian Academy of Science and Letters}}), and many exciting scientific collaborations. Janus J. Eriksen (JJE) wishes to thank Trygve for his strong, early support, indirect mentoring over the years, and continued friendship. This work was supported by two research grants awarded to JJE, no. 37411 from VILLUM FONDEN (a part of THE VELUX FOUNDATIONS) and no. 10.46540/2064-00007B from the Independent Research Fund Denmark. The authors gratefully acknowledge compute time granted on the Mogon II supercomputer, Johannes Gutenberg-Universit{\"a}t Mainz ({\url{hpc.uni-mainz.de}}).

\section*{Supporting Information}

The supporting information (SI) provides a number of additional details on the new algorithms presented in Sect. \ref{theory_sect}. Furthermore, Figs. S6--S11 report additional results for ammonia and methane on par with Figs. \ref{h2o_inc_convergence}--\ref{h2o_fit_quality}, while Fig. S12 collects detailed results behind Fig. \ref{benchmark_timings}, and Fig. S13 presents a version of Fig. \ref{entanglement_local} in a basis of standard canonical orbitals.

\section*{Data Availability}

Data in support of the findings of this study are available within the article and its SI.

\newpage

\providecommand{\latin}[1]{#1}
\makeatletter
\providecommand{\doi}
  {\begingroup\let\do\@makeother\dospecials
  \catcode`\{=1 \catcode`\}=2 \doi@aux}
\providecommand{\doi@aux}[1]{\endgroup\texttt{#1}}
\makeatother
\providecommand*\mcitethebibliography{\thebibliography}
\csname @ifundefined\endcsname{endmcitethebibliography}
  {\let\endmcitethebibliography\endthebibliography}{}


\begin{mcitethebibliography}{57}
\providecommand*\natexlab[1]{#1}
\providecommand*\mciteSetBstSublistMode[1]{}
\providecommand*\mciteSetBstMaxWidthForm[2]{}
\providecommand*\mciteBstWouldAddEndPuncttrue
  {\def\EndOfBibitem{\unskip.}}
\providecommand*\mciteBstWouldAddEndPunctfalse
  {\let\EndOfBibitem\relax}
\providecommand*\mciteSetBstMidEndSepPunct[3]{}
\providecommand*\mciteSetBstSublistLabelBeginEnd[3]{}
\providecommand*\EndOfBibitem{}
\mciteSetBstSublistMode{f}
\mciteSetBstMaxWidthForm{subitem}{(\alph{mcitesubitemcount})}
\mciteSetBstSublistLabelBeginEnd
  {\mcitemaxwidthsubitemform\space}
  {\relax}
  {\relax}

\bibitem[Eriksen(2021)]{eriksen2021a}
Eriksen,~J.~J. {The Shape of Full Configuration Interaction to Come}.
  \emph{{J}. {P}hys. {C}hem. {L}ett.} \textbf{2021}, \emph{12}, 418\relax
\mciteBstWouldAddEndPuncttrue
\mciteSetBstMidEndSepPunct{\mcitedefaultmidpunct}
{\mcitedefaultendpunct}{\mcitedefaultseppunct}\relax
\EndOfBibitem
\bibitem[Eriksen \latin{et~al.}(2017)Eriksen, Lipparini, and
  Gauss]{Eriksen2017}
Eriksen,~J.~J.; Lipparini,~F.; Gauss,~J. Virtual Orbital Many-Body Expansions:
  A Possible Route Towards the Full Configuration Interaction Limit. \emph{J.
  Phys. Chem. Lett.} \textbf{2017}, \emph{8}, 4633\relax
\mciteBstWouldAddEndPuncttrue
\mciteSetBstMidEndSepPunct{\mcitedefaultmidpunct}
{\mcitedefaultendpunct}{\mcitedefaultseppunct}\relax
\EndOfBibitem
\bibitem[Eriksen and Gauss(2018)Eriksen, and Gauss]{Eriksen2018}
Eriksen,~J.~J.; Gauss,~J. Many-Body Expanded Full Configuration Interaction. I.
  Weakly Correlated Regime. \emph{J. Chem. Theory Comput.} \textbf{2018},
  \emph{14}, 5180\relax
\mciteBstWouldAddEndPuncttrue
\mciteSetBstMidEndSepPunct{\mcitedefaultmidpunct}
{\mcitedefaultendpunct}{\mcitedefaultseppunct}\relax
\EndOfBibitem
\bibitem[Eriksen and Gauss(2019)Eriksen, and Gauss]{Eriksen2019}
Eriksen,~J.~J.; Gauss,~J. Many-Body Expanded Full Configuration Interaction.
  {II}. Strongly Correlated Regime. \emph{J. Chem. Theory Comput.}
  \textbf{2019}, \emph{15}, 4873\relax
\mciteBstWouldAddEndPuncttrue
\mciteSetBstMidEndSepPunct{\mcitedefaultmidpunct}
{\mcitedefaultendpunct}{\mcitedefaultseppunct}\relax
\EndOfBibitem
\bibitem[Eriksen and Gauss(2019)Eriksen, and Gauss]{Eriksen2019a}
Eriksen,~J.~J.; Gauss,~J. Generalized Many-Body Expanded Full Configuration
  Interaction Theory. \emph{J. Phys. Chem. Lett.} \textbf{2019}, \emph{10},
  7910\relax
\mciteBstWouldAddEndPuncttrue
\mciteSetBstMidEndSepPunct{\mcitedefaultmidpunct}
{\mcitedefaultendpunct}{\mcitedefaultseppunct}\relax
\EndOfBibitem
\bibitem[Eriksen and Gauss(2020)Eriksen, and Gauss]{eriksen2020}
Eriksen,~J.~J.; Gauss,~J. {Ground and Excited State First-Order Properties in
  Many-Body Expanded Full Configuration Interaction Theory}. \emph{{J}. {C}hem.
  {P}hys.} \textbf{2020}, \emph{153}, 154107\relax
\mciteBstWouldAddEndPuncttrue
\mciteSetBstMidEndSepPunct{\mcitedefaultmidpunct}
{\mcitedefaultendpunct}{\mcitedefaultseppunct}\relax
\EndOfBibitem
\bibitem[Eriksen and Gauss(2021)Eriksen, and Gauss]{eriksen2021}
Eriksen,~J.~J.; Gauss,~J. {Incremental Treatments of the Full Configuration
  Interaction Problem}. \emph{{W}IREs {C}omput. {M}ol. {S}ci.} \textbf{2021},
  \emph{11}, e1525\relax
\mciteBstWouldAddEndPuncttrue
\mciteSetBstMidEndSepPunct{\mcitedefaultmidpunct}
{\mcitedefaultendpunct}{\mcitedefaultseppunct}\relax
\EndOfBibitem
\bibitem[Greiner \latin{et~al.}(2024)Greiner, Gianni, Nottoli, Lipparini,
  Eriksen, and Gauss]{Greiner2024}
Greiner,~J.; Gianni,~I.; Nottoli,~T.; Lipparini,~F.; Eriksen,~J.; Gauss,~J.
  MBE-CASSCF Approach for the Accurate Treatment of Large Active Spaces.
  \emph{J. Chem. Theory Comput.} \textbf{2024}, \emph{20}, 4663\relax
\mciteBstWouldAddEndPuncttrue
\mciteSetBstMidEndSepPunct{\mcitedefaultmidpunct}
{\mcitedefaultendpunct}{\mcitedefaultseppunct}\relax
\EndOfBibitem
\bibitem[Chilkuri and Neese(2021)Chilkuri, and Neese]{neese_fci21_jctc_2021}
Chilkuri,~V.~G.; Neese,~F. {Comparison of Many-Particle Representations for
  Selected Configuration Interaction: II. Numerical Benchmark Calculations}.
  \emph{J. Chem. Theory Comput.} \textbf{2021}, \emph{17}, 2868\relax
\mciteBstWouldAddEndPuncttrue
\mciteSetBstMidEndSepPunct{\mcitedefaultmidpunct}
{\mcitedefaultendpunct}{\mcitedefaultseppunct}\relax
\EndOfBibitem
\bibitem[Uhlmann(1976)]{Uhlmann1976}
Uhlmann,~A. The ``Transition Probability" in the State Space of a *-Algebra.
  \emph{Rep. Math. Phys.} \textbf{1976}, \emph{9}, 273\relax
\mciteBstWouldAddEndPuncttrue
\mciteSetBstMidEndSepPunct{\mcitedefaultmidpunct}
{\mcitedefaultendpunct}{\mcitedefaultseppunct}\relax
\EndOfBibitem
\bibitem[Alberti(1983)]{Alberti1983}
Alberti,~P.~M. A Note on the Transition Probability over C*-Algebras.
  \emph{Lett. Math. Phys.} \textbf{1983}, \emph{7}, 25\relax
\mciteBstWouldAddEndPuncttrue
\mciteSetBstMidEndSepPunct{\mcitedefaultmidpunct}
{\mcitedefaultendpunct}{\mcitedefaultseppunct}\relax
\EndOfBibitem
\bibitem[Peres(1984)]{Peres1984}
Peres,~A. Stability of Quantum Motion in Chaotic and Regular Systems.
  \emph{Phys. Rev. A} \textbf{1984}, \emph{30}, 1610\relax
\mciteBstWouldAddEndPuncttrue
\mciteSetBstMidEndSepPunct{\mcitedefaultmidpunct}
{\mcitedefaultendpunct}{\mcitedefaultseppunct}\relax
\EndOfBibitem
\bibitem[Jozsa(1994)]{Jozsa1994}
Jozsa,~R. Fidelity for Mixed Quantum States. \emph{J. Mod. Optic.}
  \textbf{1994}, \emph{41}, 2315\relax
\mciteBstWouldAddEndPuncttrue
\mciteSetBstMidEndSepPunct{\mcitedefaultmidpunct}
{\mcitedefaultendpunct}{\mcitedefaultseppunct}\relax
\EndOfBibitem
\bibitem[Boguslawski \latin{et~al.}(2011)Boguslawski, Marti, and
  Reiher]{Boguslawski2011}
Boguslawski,~K.; Marti,~K.~H.; Reiher,~M. Construction of {CASCI}-Type Wave
  Functions for Very Large Active Spaces. \emph{J. Chem. Phys.} \textbf{2011},
  \emph{134}, 224101\relax
\mciteBstWouldAddEndPuncttrue
\mciteSetBstMidEndSepPunct{\mcitedefaultmidpunct}
{\mcitedefaultendpunct}{\mcitedefaultseppunct}\relax
\EndOfBibitem
\bibitem[Not()]{Note-1}
We note how, when calculating the quantum fidelity of two CASCI states of which
  one active space is a subspace of the other, special care must be taken to
  account for the correct sign of the individual coefficients in Eq.
  \ref{quant_fidel_eq} in cases where the particle number in the two active
  spaces differs.\relax
\mciteBstWouldAddEndPunctfalse
\mciteSetBstMidEndSepPunct{\mcitedefaultmidpunct}
{}{\mcitedefaultseppunct}\relax
\EndOfBibitem
\bibitem[{Purvis, III} and Bartlett(1982){Purvis, III}, and
  Bartlett]{ccsd_paper_1_jcp_1982}
{Purvis, III},~G.~D.; Bartlett,~R.~J. {A Full Coupled-Cluster Singles and
  Doubles Model: The Inclusion of Disconnected Triples}. \emph{{J}. {C}hem.
  {P}hys.} \textbf{1982}, \emph{76}, 1910\relax
\mciteBstWouldAddEndPuncttrue
\mciteSetBstMidEndSepPunct{\mcitedefaultmidpunct}
{\mcitedefaultendpunct}{\mcitedefaultseppunct}\relax
\EndOfBibitem
\bibitem[Raghavachari \latin{et~al.}(1989)Raghavachari, Trucks, Pople, and
  Head-Gordon]{original_ccsdpt_paper}
Raghavachari,~K.; Trucks,~G.~W.; Pople,~J.~A.; Head-Gordon,~M. {A Fifth-Order
  Perturbation Comparison of Electron Correlation Theories}. \emph{{C}hem.
  {P}hys. {L}ett.} \textbf{1989}, \emph{157}, 479\relax
\mciteBstWouldAddEndPuncttrue
\mciteSetBstMidEndSepPunct{\mcitedefaultmidpunct}
{\mcitedefaultendpunct}{\mcitedefaultseppunct}\relax
\EndOfBibitem
\bibitem[Abraham and Mayhall(2021)Abraham, and Mayhall]{Abraham2021}
Abraham,~V.; Mayhall,~N. Cluster Many-Body Expansion: A Many-Body Expansion of
  the Electron Correlation Energy about a Cluster Mean-Field Reference.
  \emph{J. Chem. Phys.} \textbf{2021}, \emph{155}, 054101\relax
\mciteBstWouldAddEndPuncttrue
\mciteSetBstMidEndSepPunct{\mcitedefaultmidpunct}
{\mcitedefaultendpunct}{\mcitedefaultseppunct}\relax
\EndOfBibitem
\bibitem[Jim{\'e}nez-Hoyos and Scuseria(2015)Jim{\'e}nez-Hoyos, and
  Scuseria]{JimenezHoyos2015}
Jim{\'e}nez-Hoyos,~C.~A.; Scuseria,~G.~E. Cluster-Based Mean-Field and
  Perturbative Description of Strongly Correlated Fermion Systems: Application
  to the One- and Two-Dimensional Hubbard Model. \emph{Phys. Rev. B}
  \textbf{2015}, \emph{92}, 085101\relax
\mciteBstWouldAddEndPuncttrue
\mciteSetBstMidEndSepPunct{\mcitedefaultmidpunct}
{\mcitedefaultendpunct}{\mcitedefaultseppunct}\relax
\EndOfBibitem
\bibitem[Zimmerman and Rask(2019)Zimmerman, and Rask]{Zimmerman2019}
Zimmerman,~P.~M.; Rask,~A.~E. Evaluation of Full Valence Correlation Energies
  and Gradients. \emph{J. Chem. Phys.} \textbf{2019}, \emph{150}, 244117\relax
\mciteBstWouldAddEndPuncttrue
\mciteSetBstMidEndSepPunct{\mcitedefaultmidpunct}
{\mcitedefaultendpunct}{\mcitedefaultseppunct}\relax
\EndOfBibitem
\bibitem[Hurley \latin{et~al.}(1953)Hurley, Lennard-Jones, and
  Pople]{Hurley1953}
Hurley,~A.~C.; Lennard-Jones,~J.~E.; Pople,~J.~A. The Molecular Orbital Theory
  of Chemical Valency {XVI}. A Theory of Paired-Electrons in Polyatomic
  Molecules. \emph{Proc. R. Soc. A} \textbf{1953}, \emph{220}, 446\relax
\mciteBstWouldAddEndPuncttrue
\mciteSetBstMidEndSepPunct{\mcitedefaultmidpunct}
{\mcitedefaultendpunct}{\mcitedefaultseppunct}\relax
\EndOfBibitem
\bibitem[Goddard and Harding(1978)Goddard, and Harding]{Goddard1978}
Goddard,~W.~A.; Harding,~L.~B. The Description of Chemical Bonding From
  {\textit{Ab Initio}} Calculations. \emph{Annu. Rev. Phys. Chem.}
  \textbf{1978}, \emph{29}, 363\relax
\mciteBstWouldAddEndPuncttrue
\mciteSetBstMidEndSepPunct{\mcitedefaultmidpunct}
{\mcitedefaultendpunct}{\mcitedefaultseppunct}\relax
\EndOfBibitem
\bibitem[Langlois \latin{et~al.}(1990)Langlois, Muller, Coley, Goddard,
  Ringnalda, Won, and Friesner]{Langlois1990}
Langlois,~J.-M.; Muller,~R.~P.; Coley,~T.~R.; Goddard,~W.~A.; Ringnalda,~M.~N.;
  Won,~Y.; Friesner,~R.~A. Pseudospectral Generalized Valence-Bond
  Calculations: Application to Methylene, Ethylene, and Silylene. \emph{J.
  Chem. Phys.} \textbf{1990}, \emph{92}, 7488\relax
\mciteBstWouldAddEndPuncttrue
\mciteSetBstMidEndSepPunct{\mcitedefaultmidpunct}
{\mcitedefaultendpunct}{\mcitedefaultseppunct}\relax
\EndOfBibitem
\bibitem[Gerratt \latin{et~al.}(1997)Gerratt, Cooper, Karadakov, and
  Raimondi]{Gerratt1997}
Gerratt,~J.; Cooper,~D.~L.; Karadakov,~P.~B.; Raimondi,~M. Modern Valence Bond
  Theory. \emph{Chem. Soc. Rev.} \textbf{1997}, \emph{26}, 87\relax
\mciteBstWouldAddEndPuncttrue
\mciteSetBstMidEndSepPunct{\mcitedefaultmidpunct}
{\mcitedefaultendpunct}{\mcitedefaultseppunct}\relax
\EndOfBibitem
\bibitem[Friedrich \latin{et~al.}(2007)Friedrich, Hanrath, and
  Dolg]{Friedrich2007}
Friedrich,~J.; Hanrath,~M.; Dolg,~M. Fully Automated Implementation of the
  Incremental Scheme: Application to {CCSD} Energies for Hydrocarbons and
  Transition Metal Compounds. \emph{J. Chem. Phys.} \textbf{2007}, \emph{126},
  154110\relax
\mciteBstWouldAddEndPuncttrue
\mciteSetBstMidEndSepPunct{\mcitedefaultmidpunct}
{\mcitedefaultendpunct}{\mcitedefaultseppunct}\relax
\EndOfBibitem
\bibitem[Goertzel(1949)]{Goertzel1949}
Goertzel,~G. Quota Sampling and Importance Functions in Stochastic Solution of
  Particle Problems. \emph{Technical Report ORNL-434, Oak Ridge National
  Laboratory} \textbf{1949}, \relax
\mciteBstWouldAddEndPunctfalse
\mciteSetBstMidEndSepPunct{\mcitedefaultmidpunct}
{}{\mcitedefaultseppunct}\relax
\EndOfBibitem
\bibitem[Kahn and Harris(1951)Kahn, and Harris]{Kahn1951}
Kahn,~H.; Harris,~T.~E. In \emph{Monte Carlo Method}; Householder,~A.~S., Ed.;
  Applied Mathematics Series; National Bureau of Standards, 1951; p~27\relax
\mciteBstWouldAddEndPuncttrue
\mciteSetBstMidEndSepPunct{\mcitedefaultmidpunct}
{\mcitedefaultendpunct}{\mcitedefaultseppunct}\relax
\EndOfBibitem
\bibitem[Rosenblatt(1956)]{Rosenblatt1956}
Rosenblatt,~M. Remarks on Some Nonparametric Estimates of a Density Function.
  \emph{Ann. Math. Stat.} \textbf{1956}, \emph{27}, 832\relax
\mciteBstWouldAddEndPuncttrue
\mciteSetBstMidEndSepPunct{\mcitedefaultmidpunct}
{\mcitedefaultendpunct}{\mcitedefaultseppunct}\relax
\EndOfBibitem
\bibitem[Parzen(1962)]{Parzen1962}
Parzen,~E. On Estimation of a Probability Density Function and Mode. \emph{Ann.
  Math. Stat.} \textbf{1962}, \emph{33}, 1065\relax
\mciteBstWouldAddEndPuncttrue
\mciteSetBstMidEndSepPunct{\mcitedefaultmidpunct}
{\mcitedefaultendpunct}{\mcitedefaultseppunct}\relax
\EndOfBibitem
\bibitem[Greiner and Eriksen()Greiner, and Eriksen]{pymbe}
Greiner,~J.; Eriksen,~J.~J. Py{MBE}: A Many-Body Expanded Correlation Code.
  See: \url{https://gitlab.com/januseriksen/pymbe}\relax
\mciteBstWouldAddEndPuncttrue
\mciteSetBstMidEndSepPunct{\mcitedefaultmidpunct}
{\mcitedefaultendpunct}{\mcitedefaultseppunct}\relax
\EndOfBibitem
\bibitem[Sun \latin{et~al.}(2017)Sun, Berkelbach, Blunt, Booth, Guo, Li, Liu,
  McClain, Sayfutyarova, Sharma, Wouters, and Chan]{Sun2017}
Sun,~Q.; Berkelbach,~T.~C.; Blunt,~N.~S.; Booth,~G.~H.; Guo,~S.; Li,~Z.;
  Liu,~J.; McClain,~J.~D.; Sayfutyarova,~E.~R.; Sharma,~S.; Wouters,~S.;
  Chan,~G. K.-L. {PySCF}: The Python-Based Simulations of Chemistry Framework.
  \emph{Wiley Interdiscip. Rev.: Comput. Mol. Sci.} \textbf{2017}, \emph{8},
  e1340\relax
\mciteBstWouldAddEndPuncttrue
\mciteSetBstMidEndSepPunct{\mcitedefaultmidpunct}
{\mcitedefaultendpunct}{\mcitedefaultseppunct}\relax
\EndOfBibitem
\bibitem[Sun \latin{et~al.}(2020)Sun, Zhang, Banerjee, Bao, Barbry, Blunt,
  Bogdanov, Booth, Chen, Cui, Eriksen, Gao, Guo, Hermann, Hermes, Koh, Koval,
  Lehtola, Li, Liu, Mardirossian, McClain, Motta, Mussard, Pham, Pulkin,
  Purwanto, Robinson, Ronca, Sayfutyarova, Scheurer, Schurkus, Smith, Sun, Sun,
  Upadhyay, Wagner, Wang, White, Whitfield, Williamson, Wouters, Yang, Yu, Zhu,
  Berkelbach, Sharma, Sokolov, and Chan]{Sun2020}
Sun,~Q.; Zhang,~X.; Banerjee,~S.; Bao,~P.; Barbry,~M.; Blunt,~N.~S.;
  Bogdanov,~N.~A.; Booth,~G.~H.; Chen,~J.; Cui,~Z.-H.; Eriksen,~J.~J.; Gao,~Y.;
  Guo,~S.; Hermann,~J.; Hermes,~M.~R.; Koh,~K.; Koval,~P.; Lehtola,~S.; Li,~Z.;
  Liu,~J.; Mardirossian,~N.; McClain,~J.~D.; Motta,~M.; Mussard,~B.;
  Pham,~H.~Q.; Pulkin,~A.; Purwanto,~W.; Robinson,~P.~J.; Ronca,~E.;
  Sayfutyarova,~E.~R.; Scheurer,~M.; Schurkus,~H.~F.; Smith,~J. E.~T.; Sun,~C.;
  Sun,~S.-N.; Upadhyay,~S.; Wagner,~L.~K.; Wang,~X.; White,~A.;
  Whitfield,~J.~D.; Williamson,~M.~J.; Wouters,~S.; Yang,~J.; Yu,~J.~M.;
  Zhu,~T.; Berkelbach,~T.~C.; Sharma,~S.; Sokolov,~A.~Y.; Chan,~G. K.-L. Recent
  Developments in the {\texttt{PySCF}} Program Package. \emph{J. Chem. Phys.}
  \textbf{2020}, \emph{153}, 024109\relax
\mciteBstWouldAddEndPuncttrue
\mciteSetBstMidEndSepPunct{\mcitedefaultmidpunct}
{\mcitedefaultendpunct}{\mcitedefaultseppunct}\relax
\EndOfBibitem
\bibitem[Matthews \latin{et~al.}(2020)Matthews, Cheng, Harding, Lipparini,
  Stopkowicz, Jagau, Szalay, Gauss, and Stanton]{Matthews2020}
Matthews,~D.~A.; Cheng,~L.; Harding,~M.~E.; Lipparini,~F.; Stopkowicz,~S.;
  Jagau,~T.-C.; Szalay,~P.~G.; Gauss,~J.; Stanton,~J.~F. Coupled-Cluster
  Techniques for Computational Chemistry: The \texttt{CFOUR} Program Package.
  \emph{J. Chem. Phys.} \textbf{2020}, \emph{152}, 214108\relax
\mciteBstWouldAddEndPuncttrue
\mciteSetBstMidEndSepPunct{\mcitedefaultmidpunct}
{\mcitedefaultendpunct}{\mcitedefaultseppunct}\relax
\EndOfBibitem
\bibitem[Stanton \latin{et~al.}()Stanton, Gauss, Cheng, Harding, Matthews, and
  Szalay]{cfour}
Stanton,~J.~F.; Gauss,~J.; Cheng,~L.; Harding,~M.~E.; Matthews,~D.~A.;
  Szalay,~P.~G. {{\texttt{CFOUR}}, Coupled-Cluster techniques for Computational
  Chemistry, a quantum-chemical program package}. {W}ith contributions from
  {A}. {A}sthana, {A}.{A}. {A}uer, {R}.{J}. {B}artlett, {U}. {B}enedikt, {C}.
  {B}erger, {D}.{E}. {B}ernholdt, {S}. {B}laschke, {Y}. {J}. {B}omble, {S}.
  {B}urger, {O}. {C}hristiansen, {D}. {D}atta, {F}. {E}ngel, {R}. {F}aber, {J}.
  {G}reiner, {M}. {H}eckert, {O}. {H}eun, {M}. Hilgenberg, {C}. {H}uber,
  {T}.-{C}. {J}agau, {D}. {J}onsson, {J}. {J}us{\'e}lius, {T}. Kirsch,
  {M}.-{P}. {K}itsaras, {K}. {K}lein, {G}.{M}. {K}opper, {W}.{J}. {L}auderdale,
  {F}. {L}ipparini, {J}. {L}iu, {T}. {M}etzroth, {L}.{A}. {M}{\"u}ck, {D}.{P}.
  {O}'{N}eill, {T}. {N}ottoli, {J}. {O}swald, {D}.{R}. {P}rice, {E}.
  {P}rochnow, {C}. {P}uzzarini, {K}. {R}uud, {F}. {S}chiffmann, {W}.
  {S}chwalbach, {C}. {S}immons, {S}. {S}topkowicz, {A}. {T}ajti, {T.} Uhlirova,
  {J}. {V}{\'a}zquez, {F}. {W}ang, {J}.{D}. {W}atts, {P.} Yerg{\"u}n, {C}.
  {Z}hang, {X}. {Z}heng, and the integral packages {\texttt{MOLECULE}} ({J}.
  {A}lml{\"o}f and {P}.{R}. {T}aylor), {\texttt{PROPS}} ({P}.{R}. {T}aylor),
  {\texttt{ABACUS}} ({T}. {H}elgaker, {H}. {J}. {A}a. {J}ensen, {P}.
  {J}{\o}rgensen, and {J}. {O}lsen), and {ECP} routines by {A}. {V}. {M}itin
  and {C}. van {W}{\"u}llen. {F}or the current version, see
  {\url{http://www.cfour.de}}.\relax
\mciteBstWouldAddEndPunctfalse
\mciteSetBstMidEndSepPunct{\mcitedefaultmidpunct}
{}{\mcitedefaultseppunct}\relax
\EndOfBibitem
\bibitem[Virtanen \latin{et~al.}(2020)Virtanen, Gommers, Oliphant, Haberland,
  Reddy, Cournapeau, Burovski, Peterson, Weckesser, Bright, van~der Walt,
  Brett, Wilson, Millman, Mayorov, Nelson, Jones, Kern, Larson, Carey, Polat,
  Feng, Moore, VanderPlas, Laxalde, Perktold, Cimrman, Henriksen, Quintero,
  Harris, Archibald, Ribeiro, Pedregosa, and van Mulbregt]{Virtanen2020}
Virtanen,~P.; Gommers,~R.; Oliphant,~T.~E.; Haberland,~M.; Reddy,~T.;
  Cournapeau,~D.; Burovski,~E.; Peterson,~P.; Weckesser,~W.; Bright,~J.;
  van~der Walt,~S.~J.; Brett,~M.; Wilson,~J.; Millman,~K.~J.; Mayorov,~N.;
  Nelson,~A. R.~J.; Jones,~E.; Kern,~R.; Larson,~E.; Carey,~C.~J.;
  Polat,~{\.{I}}.; Feng,~Y.; Moore,~E.~W.; VanderPlas,~J.; Laxalde,~D.;
  Perktold,~J.; Cimrman,~R.; Henriksen,~I.; Quintero,~E.~A.; Harris,~C.~R.;
  Archibald,~A.~M.; Ribeiro,~A.~H.; Pedregosa,~F.; van Mulbregt,~P.
  {\texttt{SciPy 1.0}}: Fundamental Algorithms for Scientific Computing in
  Python. \emph{Nat. Methods} \textbf{2020}, \emph{17}, 261\relax
\mciteBstWouldAddEndPuncttrue
\mciteSetBstMidEndSepPunct{\mcitedefaultmidpunct}
{\mcitedefaultendpunct}{\mcitedefaultseppunct}\relax
\EndOfBibitem
\bibitem[Scott(1992)]{Scott1992}
Scott,~D.~W. \emph{Multivariate Density Estimation: Theory, Practice, and
  Visualization}; Wiley, 1992\relax
\mciteBstWouldAddEndPuncttrue
\mciteSetBstMidEndSepPunct{\mcitedefaultmidpunct}
{\mcitedefaultendpunct}{\mcitedefaultseppunct}\relax
\EndOfBibitem
\bibitem[Dunning(1989)]{Dunning1989}
Dunning,~T.~H. Gaussian Basis Sets for Use in Correlated Molecular
  Calculations. I. The Atoms Boron through Neon and Hydrogen. \emph{J. Chem.
  Phys.} \textbf{1989}, \emph{90}, 1007\relax
\mciteBstWouldAddEndPuncttrue
\mciteSetBstMidEndSepPunct{\mcitedefaultmidpunct}
{\mcitedefaultendpunct}{\mcitedefaultseppunct}\relax
\EndOfBibitem
\bibitem[Weigend and Ahlrichs(2005)Weigend, and Ahlrichs]{Weigend2005}
Weigend,~F.; Ahlrichs,~R. Balanced Basis Sets of Split Valence, Triple Zeta
  Valence and Quadruple Zeta Valence Quality for H to Rn: Design and Assessment
  of Accuracy. \emph{Phys. Chem. Chem. Phys.} \textbf{2005}, \emph{7},
  3297\relax
\mciteBstWouldAddEndPuncttrue
\mciteSetBstMidEndSepPunct{\mcitedefaultmidpunct}
{\mcitedefaultendpunct}{\mcitedefaultseppunct}\relax
\EndOfBibitem
\bibitem[Pipek and Mezey(1989)Pipek, and Mezey]{Pipek1989}
Pipek,~J.; Mezey,~P.~G. A Fast Intrinsic Localization Procedure Applicable for
  {\textit{Ab Initio}} and Semiempirical Linear Combination of Atomic Orbital
  Wave Functions. \emph{J. Chem. Phys.} \textbf{1989}, \emph{90}, 4916\relax
\mciteBstWouldAddEndPuncttrue
\mciteSetBstMidEndSepPunct{\mcitedefaultmidpunct}
{\mcitedefaultendpunct}{\mcitedefaultseppunct}\relax
\EndOfBibitem
\bibitem[Foster and Boys(1960)Foster, and Boys]{Foster1960}
Foster,~J.~M.; Boys,~S.~F. Canonical Configurational Interaction Procedure.
  \emph{Rev. Mod. Phys} \textbf{1960}, \emph{32}, 300\relax
\mciteBstWouldAddEndPuncttrue
\mciteSetBstMidEndSepPunct{\mcitedefaultmidpunct}
{\mcitedefaultendpunct}{\mcitedefaultseppunct}\relax
\EndOfBibitem
\bibitem[Not()]{Note-2}
As the {\texttt{PySCF}} FCI solver is restricted to Abelian and linear point
  groups, the $A'$ state was targeted during the CASSCF optimization for the
  \ce{NH3} and \ce{CH4} systems.\relax
\mciteBstWouldAddEndPunctfalse
\mciteSetBstMidEndSepPunct{\mcitedefaultmidpunct}
{}{\mcitedefaultseppunct}\relax
\EndOfBibitem
\bibitem[Stein and Reiher(2016)Stein, and Reiher]{Stein2016}
Stein,~C.~J.; Reiher,~M. Automated Selection of Active Orbital Spaces. \emph{J.
  Chem. Theory Comput.} \textbf{2016}, \emph{12}, 1760\relax
\mciteBstWouldAddEndPuncttrue
\mciteSetBstMidEndSepPunct{\mcitedefaultmidpunct}
{\mcitedefaultendpunct}{\mcitedefaultseppunct}\relax
\EndOfBibitem
\bibitem[Stein and Reiher(2019)Stein, and Reiher]{Stein2019}
Stein,~C.~J.; Reiher,~M. {\texttt{autoCAS}}: A Program for Fully Automated
  Multiconfigurational Calculations. \emph{J. Comput. Chem.} \textbf{2019},
  \emph{40}, 2216\relax
\mciteBstWouldAddEndPuncttrue
\mciteSetBstMidEndSepPunct{\mcitedefaultmidpunct}
{\mcitedefaultendpunct}{\mcitedefaultseppunct}\relax
\EndOfBibitem
\bibitem[Zhai \latin{et~al.}(2023)Zhai, Larsson, Lee, Cui, Zhu, Sun, Peng,
  Peng, Liao, Tölle, Yang, Li, and Chan]{Zhai2023}
Zhai,~H.; Larsson,~H.~R.; Lee,~S.; Cui,~Z.-H.; Zhu,~T.; Sun,~C.; Peng,~L.;
  Peng,~R.; Liao,~K.; Tölle,~J.; Yang,~J.; Li,~S.; Chan,~G. K.-L.
  {\texttt{BLOCK2}}: A Comprehensive Open Source Framework to Develop and Apply
  State-of-the-Art DMRG Algorithms in Electronic Structure and Beyond. \emph{J.
  Chem. Phys} \textbf{2023}, \emph{159}, 234801\relax
\mciteBstWouldAddEndPuncttrue
\mciteSetBstMidEndSepPunct{\mcitedefaultmidpunct}
{\mcitedefaultendpunct}{\mcitedefaultseppunct}\relax
\EndOfBibitem
\bibitem[Keller \latin{et~al.}(2015)Keller, Boguslawski, Janowski, Reiher, and
  Pulay]{Keller2015}
Keller,~S.; Boguslawski,~K.; Janowski,~T.; Reiher,~M.; Pulay,~P. Selection of
  Active Spaces for Multiconfigurational Wavefunctions. \emph{J. Chem. Phys.}
  \textbf{2015}, \emph{142}, 244104\relax
\mciteBstWouldAddEndPuncttrue
\mciteSetBstMidEndSepPunct{\mcitedefaultmidpunct}
{\mcitedefaultendpunct}{\mcitedefaultseppunct}\relax
\EndOfBibitem
\bibitem[Not()]{Note-3}
Computer architecture: 4 Intel Xeon CPUs E5-4620 on a single node (64 threads,
  32 cores @ 2.4 GHz, 7.80 GB/thread)\relax
\mciteBstWouldAddEndPuncttrue
\mciteSetBstMidEndSepPunct{\mcitedefaultmidpunct}
{\mcitedefaultendpunct}{\mcitedefaultseppunct}\relax
\EndOfBibitem
\bibitem[Jensen \latin{et~al.}(1988)Jensen, Jørgensen, Ågren, and
  Olsen]{Jensen1988}
Jensen,~H. J.~A.; Jørgensen,~P.; Ågren,~H.; Olsen,~J. Second-Order
  Møller–Plesset Perturbation Theory as a Configuration and Orbital
  Generator in Multiconfiguration Self-Consistent Field Calculations. \emph{J.
  Chem. Phys.} \textbf{1988}, \emph{88}, 3834\relax
\mciteBstWouldAddEndPuncttrue
\mciteSetBstMidEndSepPunct{\mcitedefaultmidpunct}
{\mcitedefaultendpunct}{\mcitedefaultseppunct}\relax
\EndOfBibitem
\bibitem[Bofill and Pulay(1989)Bofill, and Pulay]{Bofill1989}
Bofill,~J.~M.; Pulay,~P. The Unrestricted Natural Orbital–Complete Active
  Space (UNO–CAS) Method: An Inexpensive Alternative to the Complete Active
  Space–Self-Consistent-Field (CAS-SCF) Method. \emph{J. Chem. Phys.}
  \textbf{1989}, \emph{90}, 3637\relax
\mciteBstWouldAddEndPuncttrue
\mciteSetBstMidEndSepPunct{\mcitedefaultmidpunct}
{\mcitedefaultendpunct}{\mcitedefaultseppunct}\relax
\EndOfBibitem
\bibitem[Bensberg and Reiher(2023)Bensberg, and Reiher]{Bensberg2023}
Bensberg,~M.; Reiher,~M. Corresponding Active Orbital Spaces Along Chemical
  Reaction Paths. \emph{J. Phys. Chem. Lett.} \textbf{2023}, \emph{14},
  2112\relax
\mciteBstWouldAddEndPuncttrue
\mciteSetBstMidEndSepPunct{\mcitedefaultmidpunct}
{\mcitedefaultendpunct}{\mcitedefaultseppunct}\relax
\EndOfBibitem
\bibitem[Sayfutyarova \latin{et~al.}(2017)Sayfutyarova, Sun, Chan, and
  Knizia]{Sayfutyarova2017}
Sayfutyarova,~E.~R.; Sun,~Q.; Chan,~G. K.-L.; Knizia,~G. Automated Construction
  of Molecular Active Spaces from Atomic Valence Orbitals. \emph{J. Chem.
  Theory Comput.} \textbf{2017}, \emph{13}, 4063\relax
\mciteBstWouldAddEndPuncttrue
\mciteSetBstMidEndSepPunct{\mcitedefaultmidpunct}
{\mcitedefaultendpunct}{\mcitedefaultseppunct}\relax
\EndOfBibitem
\bibitem[Sayfutyarova and Hammes-Schiffer(2019)Sayfutyarova, and
  Hammes-Schiffer]{Sayfutyarova2019}
Sayfutyarova,~E.~R.; Hammes-Schiffer,~S. Constructing Molecular $\pi$-Orbital
  Active Spaces for Multireference Calculations of Conjugated Systems. \emph{J.
  Chem. Theory Comput.} \textbf{2019}, \emph{15}, 1679\relax
\mciteBstWouldAddEndPuncttrue
\mciteSetBstMidEndSepPunct{\mcitedefaultmidpunct}
{\mcitedefaultendpunct}{\mcitedefaultseppunct}\relax
\EndOfBibitem
\bibitem[Khedkar and Roemelt(2019)Khedkar, and Roemelt]{Khedkar2019}
Khedkar,~A.; Roemelt,~M. Active Space Selection Based on Natural Orbital
  Occupation Numbers from $N$-Electron Valence Perturbation Theory. \emph{J.
  Chem. Theory Comput.} \textbf{2019}, \emph{15}, 3522\relax
\mciteBstWouldAddEndPuncttrue
\mciteSetBstMidEndSepPunct{\mcitedefaultmidpunct}
{\mcitedefaultendpunct}{\mcitedefaultseppunct}\relax
\EndOfBibitem
\bibitem[Stemmle and Paulus(2019)Stemmle, and Paulus]{Stemmle2019}
Stemmle,~C.; Paulus,~B. Quantification of Electron Correlation Effects: Quantum
  Information Theory {vs} Method of Increments. \emph{Int. J. Quantum Chem.}
  \textbf{2019}, \emph{119}, e26007\relax
\mciteBstWouldAddEndPuncttrue
\mciteSetBstMidEndSepPunct{\mcitedefaultmidpunct}
{\mcitedefaultendpunct}{\mcitedefaultseppunct}\relax
\EndOfBibitem
\bibitem[Richard and Herbert(2012)Richard, and Herbert]{Richard2012}
Richard,~R.~M.; Herbert,~J.~M. A Generalized Many-Body Expansion and a Unified
  View of Fragment-Based Methods in Electronic Structure Theory. \emph{J. Chem.
  Phys.} \textbf{2012}, \emph{137}, 064113\relax
\mciteBstWouldAddEndPuncttrue
\mciteSetBstMidEndSepPunct{\mcitedefaultmidpunct}
{\mcitedefaultendpunct}{\mcitedefaultseppunct}\relax
\EndOfBibitem
\bibitem[Greiner and Eriksen(2023)Greiner, and Eriksen]{Greiner2023}
Greiner,~J.; Eriksen,~J.~J. Symmetrization of Localized Molecular Orbitals.
  \emph{J. Phys. Chem. A} \textbf{2023}, \emph{127}, 3535\relax
\mciteBstWouldAddEndPuncttrue
\mciteSetBstMidEndSepPunct{\mcitedefaultmidpunct}
{\mcitedefaultendpunct}{\mcitedefaultseppunct}\relax
\EndOfBibitem
\bibitem[Cox and Oakes(2018)Cox, and Oakes]{Cox2018}
Cox,~D.; Oakes,~D. \emph{Analysis of Survival Data}; Chapman and Hall/CRC,
  2018\relax
\mciteBstWouldAddEndPuncttrue
\mciteSetBstMidEndSepPunct{\mcitedefaultmidpunct}
{\mcitedefaultendpunct}{\mcitedefaultseppunct}\relax
\EndOfBibitem
\end{mcitethebibliography}
\end{document}